\documentclass[10pt,conference,letterpaper]{IEEEtran}
\usepackage[USenglish]{babel}
\usepackage{times,amsmath}
\usepackage[pdftex]{graphicx}
   \graphicspath{{./pdf/}{./png/}{./jpeg/}{./eps/}}
   \DeclareGraphicsExtensions{.pdf,.png,.jpeg,.eps}
\usepackage{cite}
\usepackage{url}
\usepackage{fmtcount}
\usepackage{color}
\usepackage[tight,footnotesize]{subfigure}
\usepackage{multicol}
\usepackage{amssymb}
\usepackage{wasysym}
\usepackage{dblfloatfix} 
\usepackage{calc}
\usepackage{fix-cm}
\usepackage[activate={true,nocompatibility},final,tracking=true,factor=500,stretch=40,shrink=40]{microtype}

\usepackage[libertine]{newtxmath}
\usepackage{soul}
\usepackage{ulem}
\usepackage{xcolor}
\usepackage{framed}
\colorlet{shadecolor}{yellow}

\renewcommand{\d}{{\mathrm d}}

\newcommand{\sgn}{{\mathrm{sgn}}}

\newcommand{\half}{{\frac{1}{2}}}
\newcommand{\Ibl}{{\llbracket}}
\newcommand{\Ibr}{{\rrbracket}}
\renewcommand{\mod}{{\mathrm{mod}}}

\DeclareMathOperator\erfc{erfc}
\newcommand{\appropto}{\mathrel{\vcenter{
  \offinterlineskip\halign{\hfil$##$\cr
    \propto\cr\noalign{\kern2pt}\sim\cr\noalign{\kern-2pt}}}}}
\newcommand{\CalphaPM}{{{\mathcal{C}}^{\boldmath \alpha^+}_{\boldmath \alpha^-}}}
\newcommand{\Seps}{{{\mathcal{S}}_{\scriptscriptstyle \varepsilon}}}
\definecolor{DarkRed}{rgb}{0.5,0,0}
\definecolor{DarkGreen}{rgb}{0,0.5,0}
\definecolor{DarkerGreen}{rgb}{0,0.3333,0}
\definecolor{DarkBlue}{rgb}{0,0,0.75}
\definecolor{RoyalBlue}{rgb}{0,0.1373,0.4000}
\definecolor{NavyBlue}{rgb}{0,0,0.5020}
\definecolor{CobaltBlue}{rgb}{0,0.2784,0.6706}
\definecolor{lightlightgray}{rgb}{0.96875,0.96875,0.96875}
\definecolor{cyan}{rgb}{0,1,1}
\newcommand{\beginlabel}[2]{%
\begin{#1}\label{#2}}
\begin{document}
\pagestyle{plain}
\title{Pulsed Waveforms and Intermittently Nonlinear Filtering in Synthesis of Low-SNR and Covert Communications}
\author{\IEEEauthorblockN{Alexei V. Nikitin}
\IEEEauthorblockA{
Nonlinear LLC\\
Wamego, Kansas, USA\\
E-mail: avn@nonlinearcorp.com}
\and
\IEEEauthorblockN{Ruslan L. Davidchack}
\IEEEauthorblockA{School of Mathematics and Actuarial Science\\ 
U. of Leicester, Leicester, UK\\
E-mail: rld8@leicester.ac.uk}}
\maketitle
\begin{abstract}
In traditional spread-spectrum techniques, a wideband transmit signal is obtained by modulating a wideband carrier by a narrowband signal containing a relatively low-rate message. In the receiver, the respective demodulation/despreading restores the information-carrying narrowband signal. In this paper, we introduce an alternative approach, where the low-rate information is encoded directly into a wideband waveform of a given bandwidth, without physical ``spreading" of the carrier's frequency. The main advantages of this approach lie in extended options for encoding the information, and in retaining a reversible control over the temporal and amplitude structures of the modulating wideband waveforms. Significant ``excess bandwidth" (over that needed to carry the information) enables us to use allpass filters to manage statistical properties and time-domain appearances of these waveforms without changing their spectral composition. For example, a mixture of transmitted waveforms can be shaped as a low-crest-factor signal (e.g. to reduce the burden on the power amplifier), and/or made statistically indistinguishable from Gaussian noise (e.g. for covert transmissions and physical layer steganography), while the selected components of the received waveform can be transformed into high-crest-factor pulse trains suitable for multiplexing and/or low-SNR communications. Further, control over the temporal and amplitude structures of wideband waveforms carrying low-rate information enables effective use of nonlinear filtering techniques. Such techniques can be employed for robust real-time asynchronous extraction of the information, as well as for separation of wideband signal components with identical spectral content from each other. This can facilitate development of a large variety of low-SNR and covert communication configurations.
\end{abstract}
\maketitle
\section{Introduction} \label{sec:introduction}
The additive white Gaussian noise (AWGN) capacity~$C$ of a channel operating in the power-limited regime (i.e. when the received signal-to-noise ratio (SNR) is small, $\mathrm{SNR} \ll 0\,$dB) can be expressed as $C\approx \bar{P}/(N_0\,\ln 2)$, where~$\bar{P}$ is the average received power and~$N_0$ is the power spectral density (PSD) of the noise~\cite{Verdu98fifty}. This capacity is linear in power and insensitive to bandwidth and, therefore, by spreading the average transmitted power of the information-carrying signal over a large frequency band, the average PSD of the signal could be made much smaller than the PSD of the noise. This would ``hide" the signal in the channel noise, making the transmission covert and insensitive to narrowband interference.

Various techniques to achieve such ``spreading" are commonly referred to as {\it spread-spectrum\/}. These include such well known approaches as frequency-hopping spread spectrum (FHSS) and direct-sequence spread spectrum (DSSS)~\cite{Simon94spreadspectrum, Torrieri18principles}, as well as chirp spread spectrum (CSS)~\cite{Blunt16overview, Hosseini2019nonlinear}. Within these techniques, a narrowband signal in the transmitter modulates a carrier that spans a wide frequency range. In the receiver, the respective demodulation, combined with despreading, is used to produce the information-carrying narrowband signal. Thus, even though the total SNR of the wideband transmitted signal can be low, obtaining the information about the carrier enables us to detect the presence of a narrowband signal through spectral measurements. For example, the FHSS is widely used in legacy military equipment for low-probability-of-intercept (LPI) communications. However, using frequency hopping for covert communications is nearly obsolete today, since modern wideband software-defined radio (SDR) receivers can capture all of the hops and put them back together (J.~E.~Gilley, personal communication, Feb. 9, 2020).

In DSSS, the narrow-band information-carrying signal of a given power is modulated by a wider-band, unit-power pseudorandom signal known as a spreading sequence. After demodulation/despreading in the receiver, the original information-carrying signal is restored. However, such demodulation requires precise synchronization, which is perhaps the most difficult and expensive aspect of a DSSS receiver design. Also, while despreading cannot be performed without the knowledge of the spreading sequence by the receiver, the spreading code by itself may not be usable to secure the channel. For example, linear spreading codes are easily decipherable once a short sequential set of chips from the sequence is known. To improve security, it would be desirable to perform a ``code hopping" in a manner akin to the frequency hopping. However, synchronization can be an extremely slow process for pseudorandom sequences, especially for large spreading waveforms (long codes), and thus such DSSS code hopping may be difficult to realize in practice.

\begin{figure*}[!t]
\centering{\includegraphics[width=17.6cm]{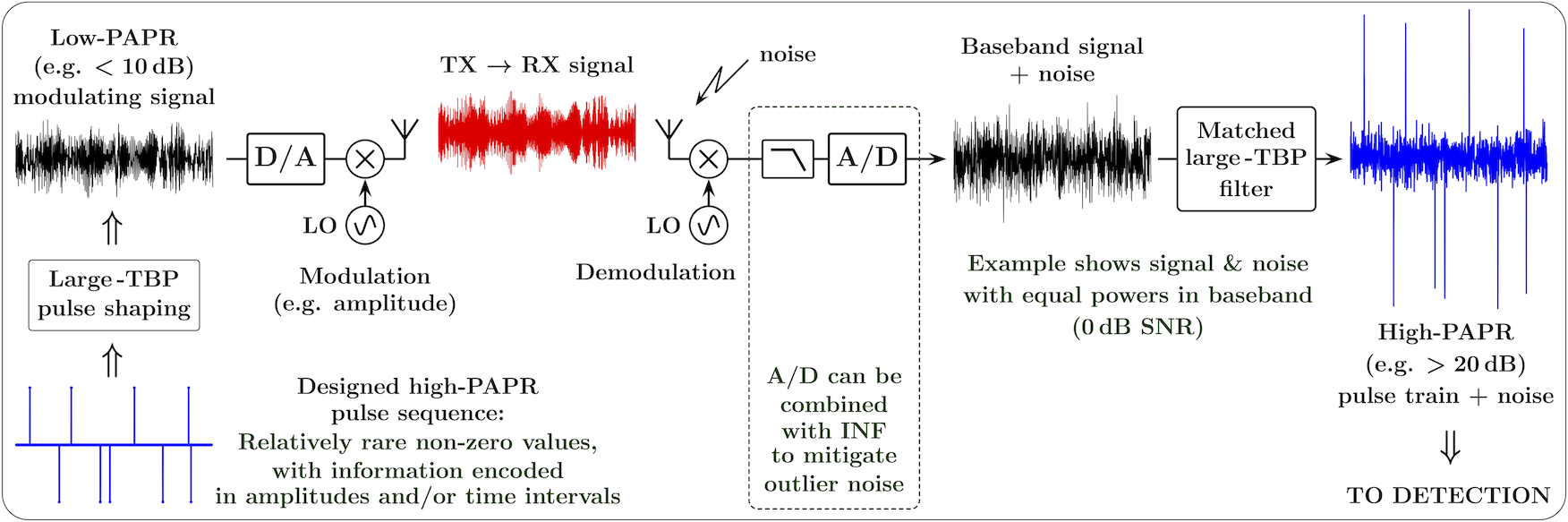}}
\caption{{\bf Using pulse trains for low-SNR communications:~} Large-TBP pulse shaping (i)~``hides" pulse train, obscuring its temporal and amplitude structure, and (ii)~reduces its PAPR, making signal suitable for transmission. In receiver, pulse train is restored by matched large-TBP filtering. High PAPR of restored pulse train enables low-SNR messaging. To make link more robust to outlier interference and to increase apparent SNR, analog-to-digital conversion in receiver can be combined with intermittently nonlinear filtering.
\label{fig:low SNR}}
\end{figure*}

In the power-limited regime, we would normally use binary coding and modulation (e.g. binary phase-shift keying (BPSK) or quadrature phase-shift keying (QPSK)) for the narrow-band information-carrying signal, and this signal will be significantly oversampled to enable wideband spreading. Thus an idealized narrow-band information-carrying signal that is to be ``spread" can be viewed as a discrete-level signal that is a linear combination of analog Heaviside unit step functions~\cite{Bracewell2000Fourier} delayed by multiples of the bit duration. Such a signal would have a limited bandwidth and a finite power. Since the derivative of the Heaviside unit step function is the Dirac $\delta$-function~\cite{Dirac58principles}, the derivative of this idealized signal will be a ``pulse train" that is a linear combination of Dirac $\delta$-functions. This pulse train will contain all the information encoded in the discrete-level signal, but it will have infinitely wide bandwidth and infinitely large power. Both the bandwidth and the power can then be reduced to the desired levels by filtering the pulse train with a lowpass or bandpass filter. As discussed in~Section~\ref{sec:ppileup}, if the time-bandwidth product (TBP) of the filter is sufficiently small so that the pulses in the filtered pulse train do not overlap, these pulses will still contain all the intended information. Such an approach is the basis for various ``impulse," ``carrier-free," and ``baseband" communication and radar systems which are collectively referred to as ``ultra wideband" (UWB)\,\cite{Barrett01history}.

Relaxing the current technical (and mostly regulatory) limitations imposed on the term ``UWB" (e.g., that the emitted signal bandwidth exceeds the lesser of 500~MHz or 20\% of the arithmetic center frequency\,\cite{ITU-RSM.1755-0}), the UWB concept can be extended to a pulsed signal in any provided frequency band. For instance, a baseband pulse train confined to a given physical band can be used for modulation of a single-frequency carrier. Then the relative bandwidth of the resulting transmit signal can be made arbitrary ``wide" (e.g. larger than $20\%$ of the carrier frequency) or ``narrow" (e.g., $<\!1\%$). If the pulse repetition rate in such a train is much smaller than its bandwidth, then such an approach can be considered a ``spread-spectrum technique."

On the one hand, transmitting low-rate information by a wideband pulsed waveform (pulse train) has an apparent appeal of no need for despreading: After demodulation, one can simply ``capture" the pulses (e.g. their amplitudes, polarities, and/or interarrival times) to obtain the encoded information. On the other hand, at first glance such a pulse train is not suitable for use as a modulating signal in practical communication systems, especially for covert communications. Indeed, let us consider a pulse train with a given average pulse rate and power. The average PSD of this train can be made arbitrary small, since it is inversely proportional to the bandwidth. However, the peak-to-average power ratio (PAPR) of such a train would be proportional to the bandwidth, making the wideband signal extremely impulsive (super-Gaussian). This leads to several considerable challenges in adapting such a signal to covert transmissions. Firstly, a high crest factor of the pulse train can put a serious burden on the transmitter hardware, potentially making this burden prohibitive (e.g. for $\mathrm{PAPR}>30\,$dB). Secondly, the high-PAPR structure of this waveform makes it easily detectable in the time domain by a large variety of techniques\,\cite{Barrett01history}, even at very low signal-to-noise ratios (SNRs), seemingly making it unsuitable for covert communications. Thirdly, it may appear that sharing the wideband channel by multiple users would require explicit allocation of the transmit and/or pulse arrival times for each sub-channel, which would be impractical in most cases.

Favorably, the temporal and amplitude structure of a wideband pulse train is modifiable by linear filtering, and such filtering can convert a high-PAPR train into a low-PAPR signal, and {\it vice versa\/}. Therefore, as detailed in this paper, such PAPR-modifying filtering enables us to use pulsed waveforms for low-SNR covert communications. Fig.~\ref{fig:low SNR} provides a simplified illustration of such an approach. The designed digital pulse sequence is a ``pulse train"~$\hat{x}[k]$ with only some of the samples having non-zero values:
\beginlabel{equation}{eq:ptrain}
  \hat{x}[k] = \sum_j \Ibl k\!=\!k_j\Ibr\, A_j\,,
\end{equation}
where $k_j$ is the sample index of the $j$-th pulse, $A_j$ is its amplitude, and the double square brackets denote the {\it Iverson bracket\/}~\cite{Knuth92two}
\beginlabel{equation}{eq:Iverson bracket}
  \Ibl P\Ibr  = \left\{
  \begin{array}{cc}
    \!\! 1 & \mathrm{if} \; P \; \mbox{is true}\\
    \!\! 0 & \mathrm{otherwise}
  \end{array}\right.,
\end{equation}
where $P$ is a statement that can be true or false. The amplitudes of the pulses~$A_j$ in such a pulse train, and/or the time intervals ${k_j\!-\!k_{j\!-\!1}}$ between the pulses, can encode the intended information. For example, the ``equidistant" train
\beginlabel{equation}{eq:ptrain equidistant}
  \hat{x}[k] = \sum_j \Ibl k\!=\!jN\Ibr\, (-1)^{b_j},
\end{equation}
where $N$ is the distance between pulses and $b_j$ is either ``0" or ``1," can encode the binary sequence~$(b_1b_2\dots b_j\dots)$. The PAPR of such a pulse train is equal to~$N$, ${\mathrm{PAPR}=N}$, and it will be high when only a small fraction of the samples has non-zero values (i.e. $N\gg 1$).

We can ``re-shape" the designed pulse train~$\hat{x}[k]$ by linear filtering:
\beginlabel{equation}{eq:ptrain filtered}
   x[k] = (\hat{x}\ast w)[k] = \sum_j A_j\, w[k\!-\!k_j]\,,
\end{equation}
where $w[k]$ is the impulse response of the filter and the asterisk denotes convolution. The filter~$w[k]$ can be, for example, a lowpass filter with a given bandwidth (smaller or equal to the Nyquist rate of the designed digital pulse sequence). As discussed in Section~\ref{sec:ppileup}, when this filter has a sufficiently large TBP, most of the samples in the reshaped train~$x[k]$ will have non-zero values, and~$x[k]$ will have a much smaller PAPR than the designed sequence~$\hat{x}[k]$. As illustrated in Fig.~\ref{fig:low SNR}, in addition to reducing the PAPR of the signal and making it suitable for modulating the carrier, large-TBP pulse shaping in the transmitter can ``hide" the pulse train, obscuring its temporal and amplitude structure.

In contrast to other modulation techniques, large-TBP pulse shaping does not attempt to avoid intersymbol interference (ISI). As we demonstrate throughout the paper, quite the opposite is true: Intentionally increasing the ``interpulse interference" in the modulating waveform carries multiple utilities. In addition to enhancing security of communications and enabling various ``layered" and multi-user configurations, these include relaxed power amplifier requirements and increased resistance to non-Gaussian noise. As the time duration of the pulse shaping filter extends over multiple interpulse intervals, the instantaneous amplitudes and/or phases of the resulting waveform\,\cite{Picinbono97instantaneous} are no longer representative of individual pulses. Instead, they are a ``piled-up" aggregate of the contributions from multiple ``stretched" pulses. Thus modulation with such a waveform may be referred to as Aggregate Spread Pulse Modulation (ASPM).

In the receiver, the demodulated signal is filtered by a large-TBP filter matched to the pulse shaping filter in the transmitter. Such filtering restores the distinct high-PAPR structure of the pulse train, without respective increase in the PAPR of the uncorrelated noise, facilitating  detection of pulses even at low SNRs. For example, in Fig.~\ref{fig:low SNR} the received signal and noise powers are equal to each other ($0$\,dB SNR in baseband). However, after the matched filtering the temporal and amplitude structure of the ``noisy" high-PAPR pulse train becomes clearly apparent and comparable with that of the designed ``ideal" pulse train. Further, to make this link more robust to outlier interference and to increase the baseband SNR, analog-to-digital conversion in the receiver can be combined with intermittently nonlinear filtering (INF) described in~\cite{Nikitin19hidden, Nikitin19complementary} and in Section~\ref{sec:INF}.

\begin{figure}[!b]
\centering{\includegraphics[width=8.6cm]{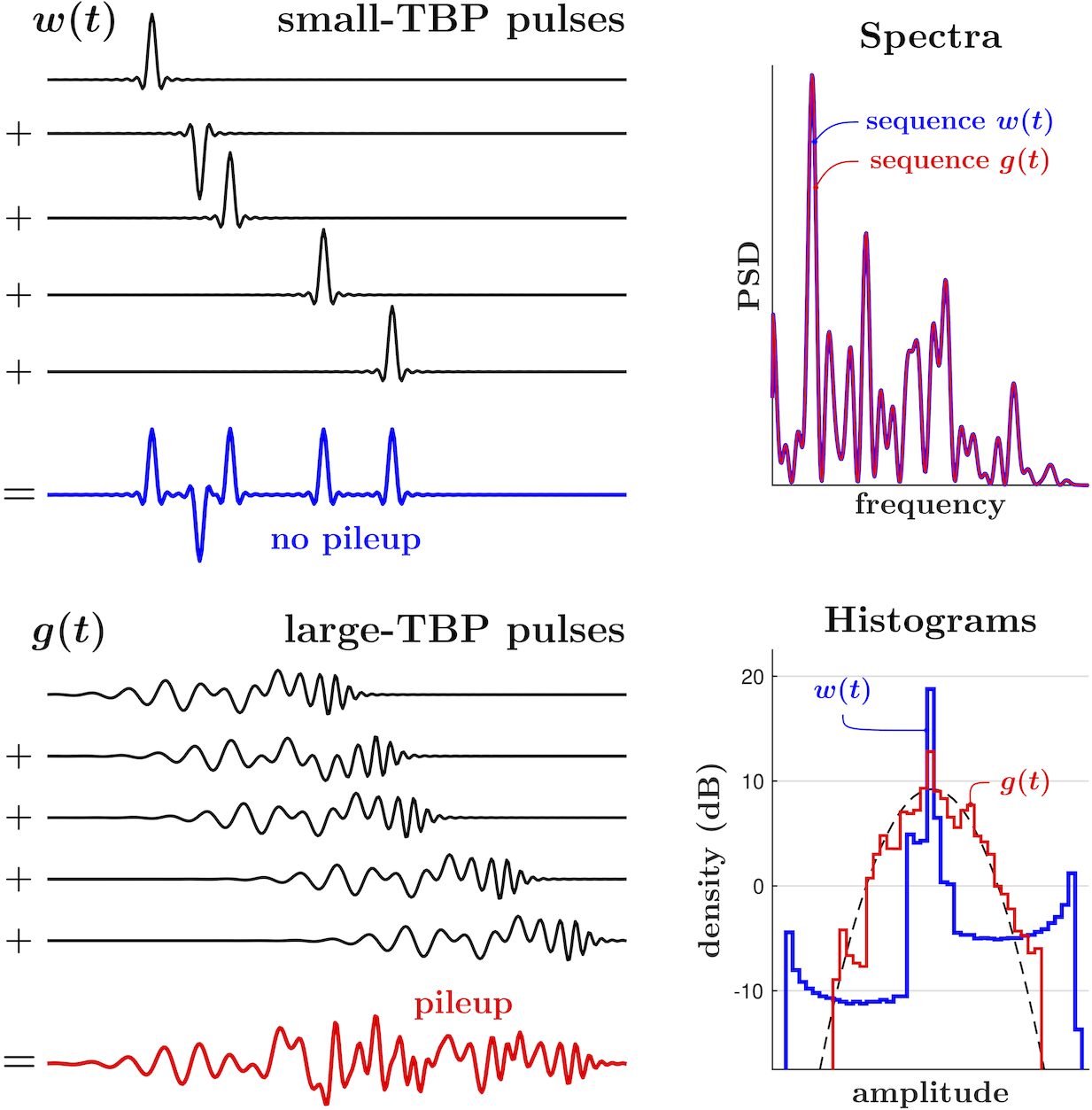}}
\caption{{\bf Illustration of pileup effect:~}%
When ``width" of pulses becomes greater than distance between them, pulses begin to overlap and interfere with each other. For pulses with same spectral content, PSDs of pulse sequences are identical, yet their temporal and amplitude structures are substantially different.
\label{fig:PPileup}}
\end{figure}

Subsequently, we interchangeably employ continuous-time (analog) and discrete (digital) representations for time-varying quantities. We use the analog representation of a signal~$x(t)$ when there are no explicit constraints on its bandwidth. When a discrete (digital) representation~$x[k]$ is used, it is assumed that~$x(t)$ is band-limited, and it is appropriately sampled so that~$x(t)$ is completely determined by~$x[k]$. Throughout the paper, while keeping some parts of the presentation rather abbreviated, we attempted to provide sufficient amount of detail required for further practical development of this approach.

\section{Scrambling and PAPR Control Utility of Pileup Effect} \label{sec:ppileup}
A pulse train~$x(t)$ can be viewed as a sum of pulses with the same shape (impulse response)~$w(t)$, same or different amplitudes~$A_j$, and distinct arrival times~$t_j$: ${x(t) = \sum_j A_j w(t\!-\!t_j)}$. When the width of the pulses in a train becomes greater than the distance between them, the pulses begin to overlap and interfere with each other. This is illustrated in Fig.~\ref{fig:PPileup}: For the same interarrival times, the pulses in the sequence consisting of the narrow pulses~$w(t)$ remain separate, while the wider (more ``spread out") pulses~$g(t)$ are ``piling up on top of each other." In this example, $g(t)$~is obtained by filtering~$w(t)$ with an allpass filter (consisting of 6~cascaded biquad sections), and thus the PSDs of the pulse sequences are identical. However, the ``pileup effect" causes the temporal and amplitude structures of these sequences to be substantially different. For a random pulse train, when the ratio of the bandwidth and the pulse arrival rate becomes significantly smaller than the TBP of a pulse, the pileup effect causes the resulting signal to become effectively Gaussian~\cite[e.g.]{Nikitin98ppileup}, making it impossible to distinguish between the individual pulses.

Indeed, let~$\hat{x}(t)$ be an ``ideal" pulse train: $\hat{x}(t) = \sum_j A_j \delta(t\!-\!t_j)$, where $\delta(x)$ is the Dirac $\delta$-function~\cite{Dirac58principles}. The {\it moving average\/} of this ideal train in a boxcar window of width~$2T$ can be represented by the convolution integral
\begin{equation} \label{eq:moving average}
  \overline{x}(t) = \int_{-\infty}^\infty \!\!\d{s}\, \frac{\theta(t\!+\!T)-\theta(t\!-\!T)}{2T}\, \hat{x}(t\!-\!s)\,,
\end{equation}
where $\theta(x)$ is the Heaviside unit step function~\cite{Bracewell2000Fourier}. At any given time~$t_i$, the value of $\overline{x}(t_i)$ is proportional to the sum of~$A_j$ for the pulses that occur within the interval~$[t_i\!-\!T,t_i\!+\!T]$. Then, if the amplitudes $A_j$ and/or the interarrival times~$t_{j+1}-t_j$ are independent and identically distributed (i.i.d.) random variables with finite mean and variance, it follows from the Central Limit Theorem~\cite[e.g.]{Aleksandrov56mathematics99} that the distribution of $\overline{x}(t_i)$ approaches Gaussian for a sufficiently large interval~$[-T,T]$.

If we replace the boxcar weighting function in~(\ref{eq:moving average}) with an arbitrary moving window~$w(t)$, then~(\ref{eq:moving average}) becomes a {\it weighted\/} moving average
\begin{equation} \label{eq:weighted moving average}
  x(t) = \int_{-\infty}^\infty \!\!\d{s}\, w(t)\, \hat{x}(t\!-\!s) = (\hat{x}\!\ast\!w)(t) = \sum_j A_j w(t\!-\!t_j)\,,
\end{equation}
which is a ``real" pulse train with the impulse response~$w(t)$. If~$w(t)$ is normalized so that~$\int_{-\infty}^\infty \d{s}\, w(s) \!=\! 1$, $w(t)$ is an {\it averaging\/} (i.e. lowpass) filter. Then, if~$w(t)$ has both the bandwidth and the TBP similar to that of the boxcar pulse of width~$2T$, the distribution of $x(t_i)$ would be similar to that of~$\overline{x}(t_i)$ (e.g. Gaussian for a sufficiently large~$T$).

\addtocounter{figure}{1}
\begin{figure*}[!b]
\centering{\includegraphics[width=14cm]{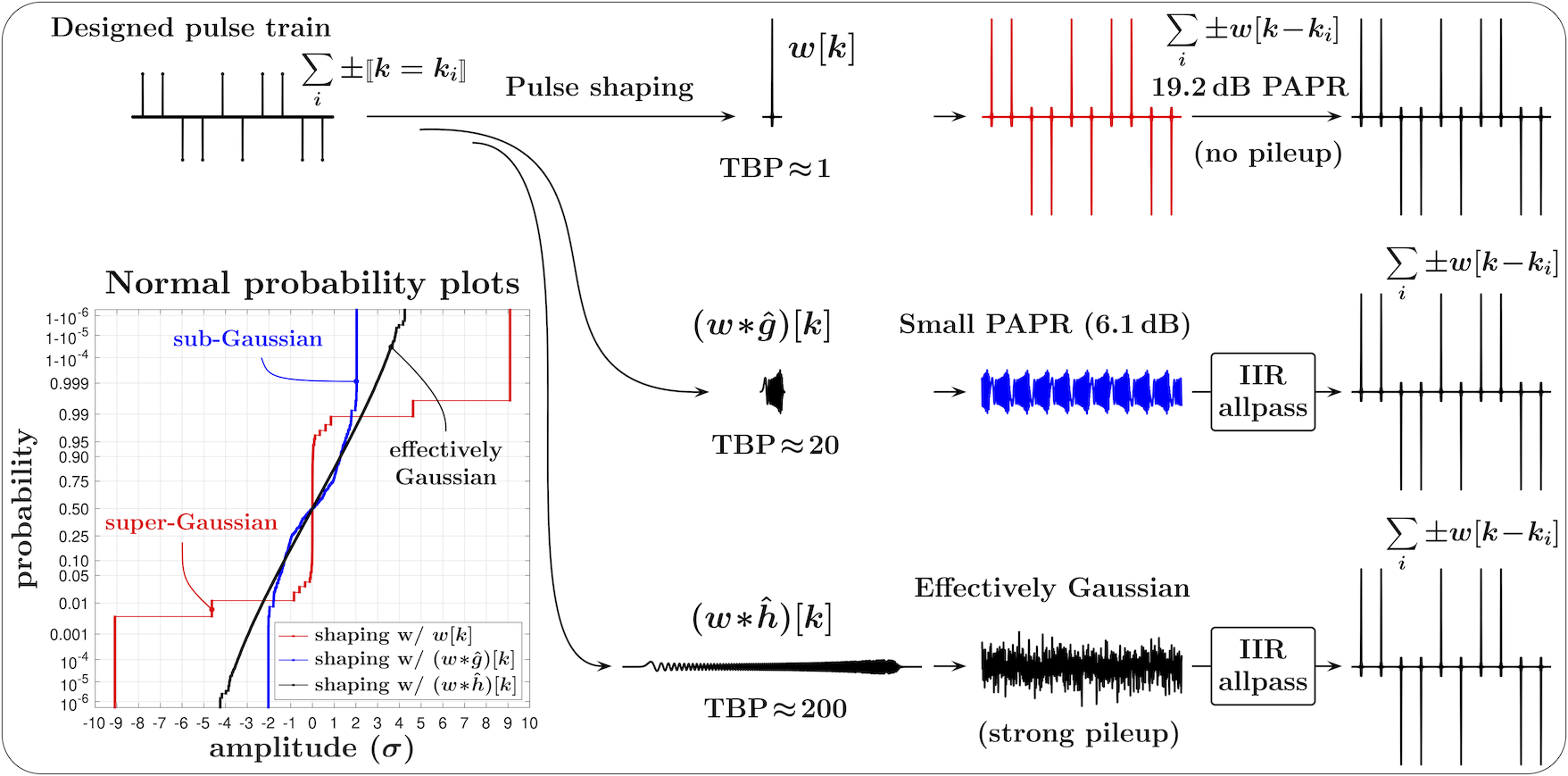}}
\caption{Using large-TBP filtering and pileup effect for PAPR control and obfuscation of temporal and amplitude structure of pulsed waveforms. In transmitter, pulse shaping with large-TBP filter reduces crest factor of pulse train, making it appear sub-Gaussian or effectively Gaussian. In receiver, signal's distinct temporal and amplitude structure is restored.
\label{fig:ppileup}}
\end{figure*}

\subsection{PAPR Control by Large-TBP Pulse Shaping} \label{subsec:PAPR control}
There are various ways to define the ``time duration" and the ``bandwidth" of a pulse~\cite[e.g.]{Barrett01history}. This can lead to a significant ambiguity in the definitions of the time-bandwidth products (TBPs), especially for waveforms with complicated temporal structures and/or frequency responses. For example, while compact support cannot be simultaneously achieved for the temporal and the spectral power densities of any pulse, the standard deviations, $\sigma_t$ and $\sigma_f$, of these power densities can be used as measures of their width\,\cite{Gabor45theory, Vetterli95wavelets}. Then, e.g., the TBP of a pulse can be defined as ${\mathrm{TBP}=4\pi\sigma_t \sigma_f \ge 1}$, with the equality (the smallest ${\mathrm{TBP}=1}$) achieved for a Gaussian pulse. However, in the context of a PAPR control function of the pileup effect, our main concern is that the change in the TBP occurs only due to the change in the temporal structure of a filter, without the respective change in its spectral content. In this case, a change in the PAPR of a pulse is indicative of the change in its ``sharpness" (or ``resolution") in the time domain, and the reciprocal of the PAPR can serve as a measure of the time duration of the pulse.

For a single pulse~$w(t)$, its PAPR can be expressed as
\begin{equation} \label{eq:PAPR}
  \mathrm{PAPR}_w = \frac{\max\left(w^2(t)\right)}{\frac{1}{T_2-T_1} \int_{T_1}^{T_2} \d{t}\, w^2(t)}\,,
\end{equation}
where the interval~$[T_1,T_2]$ includes the effective time support of~$w(t)$. Then for filters with the same spectral content but different impulse responses~$w(t)$ and~$g(t)$, the ratio of their TBPs can be expressed as the reciprocal of the ratio of their PAPRs,
\begin{equation} \label{eq:TBP}
  \frac{\mathrm{TBP}_g}{\mathrm{TBP}_w} = \frac{\max\left(w^2(t)\right)}{\max\left(g^2(t)\right)} = \frac{\mathrm{PAPR}_w}{\mathrm{PAPR}_g}\,,
\end{equation}
where the PAPRs are calculated over a sufficiently long time interval~$[T_1,T_2]$ that includes the effective time support of both filters. Note that from~(\ref{eq:TBP}) it follows that, among all possible pulses with the same spectral content, the one with the smallest TBP will contain a dominant large-magnitude peak. Hence any reasonable definition of a finite TBP for a particular filter with a given frequency response allows us to obtain comparable numerical values for the TBPs of all other filters with the same frequency response, regardless of their temporal structures.

There are multiple ways to construct pulses with identical spectral compositions (and thus identical autocorrelation functions) yet significantly different TBPs. For example, the autocorrelation function of an impulse response of any allpass filter is the Dirac $\delta$-function. Therefore, given a ``seed" small-TBP pulse with finite (FIR) or infinite (IIR) impulse response~$w(t)$, a large-TBP pulse with the same spectral content can be ``grown" from~$w(t)$ by applying a sequence of IIR allpass filters that leave the PSD of the seed pulse unmodified~\cite[e.g.]{Regalia88digital}. Then an FIR filter for pulse shaping in the transmitter can be obtained by (i)~``spreading"~$w(t)$ with an IIR allpass filter, (ii)~truncating the pulse when it sufficiently decays to zero, and (iii)~time-inverting the resulting waveform. Applying the same sequence of IIR allpass filters in the receiver to this waveform will produce the matched filter $w(-t)$ to the original seed pulse.

In the example of Fig.~\ref{fig:allpass2chirp}, the transmitter waveform is a ``piled-up" sum of thus constructed large-TBP pulses (obtained by ``spreading" $w(t)$ with the allpass filter $\hat{g}(t)$), scaled and time-shifted. In the receiver, the IIR allpass filter $g(t)$ (here consisting of 21~cascaded biquad allpass sections) recovers the underlying high-PAPR pulse train. Fig.~\ref{fig:ppileup} further illustrates how the pileup effect can be used to obscure (e.g. to mimic as Gaussian or sub-Gaussian) a large-PAPR (super-Gaussian) transmitted signal, while fully recovering its distinct temporal and amplitude structure in the receiver. In this example, pulse shaping with large-TBP filters $w\!\ast\!\hat{g}$ and $w\!\ast\!\hat{h}$ in the transmitter ``hides" the original structure of the pulse train, and the pulses with larger TBPs perform this more effectively. This can be seen in Fig.~\ref{fig:ppileup} from both the time-domain traces and the normal probability plots shown in the lower left corner. For a sufficiently large TBP, the distribution of the filtered pulse train with random pulse polarities becomes effectively Gaussian, making it impossible to distinguish between the individual pulses.

\addtocounter{figure}{-2}
\begin{figure}[!t]
\centering{\includegraphics[width=8.6cm]{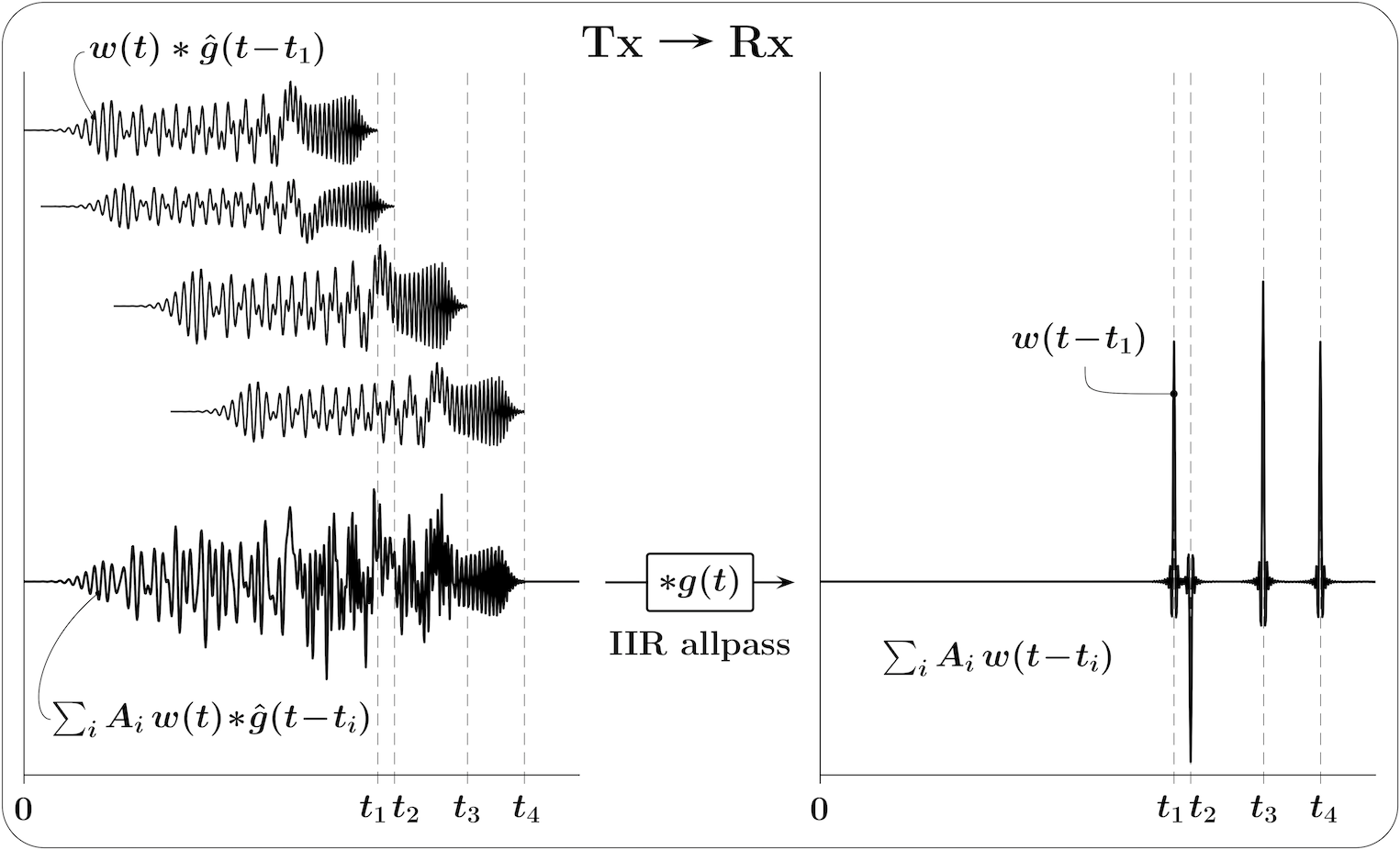}}
\caption{Transmitter waveform is constructed as sum of scaled and time-shifted large-TBP pulses. In receiver, IIR allpass filter recovers small-TBP pulse train.
\label{fig:allpass2chirp}}
\end{figure}

The seed~$w(t)$ used in Figs.~\ref{fig:allpass2chirp} and~\ref{fig:ppileup} is an FIR root-raised-cosine (RRC) filter, and thus $(w\!\ast\! w)(t)$ is a raised-cosine (RC) pulse\,\cite[e.g]{Proakis06digital}. While TBPs of RC pulses are generally larger than those of a Gaussian or a Bessel pulse, compact frequency support of RC filters is appealing for communication applications. The TBPs of these filters remain relatively small for large roll-off factors (e.g., $\mathrm{TBP} \lesssim 1.5$ for $1/3\lesssim\beta \le 1$), and in the subsequent simulations and numerical examples we use FIR RC pulses with the roll-off factor~$\beta\!=\!1/2$ ($\mathrm{TBP} \approx 1.27$).

\section{Pulsed Waveforms for Low-SNR and Covert Communications} \label{sec:pulsed waveforms}
Let us consider a pulse train consisting of pulses with a given TBP and bandwidth~$\Delta{B}$, and with the average pulse arrival rate~$\mathcal{R}$. When~$\mathcal{R}$ is sufficiently low, e.g. ${\mathcal{R}\ll\Delta{B}/\mathrm{TBP}}$, pileup is negligible. As follows from the discussion in Section~\ref{sec:ppileup}, the PAPR of such a pulse train will be inversely proportional to the pulse rate, ${\mathrm{PAPR}\propto\mathcal{R}^{-1}}$. Then, for a pulse train with a given bandwidth and average power, by reducing the pulse arrival rate the pulses can be made arbitrarily large and easily detectible even at low SNR. On the other hand, for a small PAPR and/or obfuscation of the temporal and amplitude structure of a pulsed waveform, pileup needs to be sufficiently strong. For that, for a given rate~${\mathcal{R}}$ and bandwidth~$\Delta{B}$, the TBP of the pulses needs to be sufficiently large, e.g. ${\mathrm{TBP}\gtrsim \Delta{B}/\mathcal{R}}$.

For the designed pulse train ${\hat{x}(t) = \sum_j A_j\delta(t\!-\!t_j)}$, the waveforms ${x_w(t) = \sum_j A_j w(t\!-\!t_j)}$ and ${x_g(t) = \sum_j A_j g(t\!-\!t_j)}$ can be obtained by filtering~$\hat{x}(t)$ with filters having different impulse responses~$w(t)$ and~$g(t)$, but the same frequency response and bandwidth~$\Delta{B}$. Then $x_w(t)$ would be a high-PAPR waveform suitable for messaging at a given ${\mathrm{SNR}<0\,}$dB, while~$x_g(t)$ would be a low-PAPR or effectively Gaussian ``covert" waveform, when
\beginlabel{equation}{eq:lowSNR and covert}
  \mathrm{TBP}_w\, \frac{\mathcal{R}}{\Delta{B}} \ll \mathrm{SNR}<1 \lesssim \mathrm{TBP}_g\, \frac{\mathcal{R}}{\Delta{B}}\,.
\end{equation}\\[-1.5ex]
Thus using a link with a given bandwidth~$\Delta{B}$ for covert low-SNR communications would be mainly limited by the practically obtainable value of\,~$\mathrm{TBP}_g$. For example, for~$\mathrm{TBP}_w$ of order unity and ${\mathrm{SNR} = -20\,}$dB, the ratio ${\mathcal{R}/\Delta{B}}$ would need to be smaller than approximately~$-30\,$dB and, consequently, ${\mathrm{TBP}_g\gtrsim 30\,}$dB.

For effective use of large-TBP pulse shaping for conversion of a high-PAPR pulse train with a distinct, super-Gaussian temporal and amplitude structure into an effectively Gaussian signal, the pulse train needs to be randomized. This can be accomplished by randomizing the amplitude of the pulses in the train, their arrival times, or both. The ways in which the pulse train is randomized affect the ways in which the information can be encoded and retrieved. For example, if the timing structure of the pulse train is known, synchronous pulse detection can be used. Otherwise, one may need to employ asynchronous pulse detection (e.g. the pulse counting discussed in Section~\ref{sec:INF}).

\subsection{Synchronous Pulse Detection} \label{subsec:synchronous}
Let us consider a pulse train consisting of pulses with the bandwidth~$\Delta{B}$ and a small TBP, so that a single large-magnitude peak in a pulse dominates, and assume that the arrival rate~${\mathcal{R}}$ of the pulses is sufficiently small so that pileup is negligible (e.g. ${\mathcal{R}}\ll{\mathcal{R}}_0\!=\!\half\Delta{B}/\mathrm{TBP}$). When the arrival time of a pulse with the peak magnitude~$|A|$ is known, the probability of correctly detecting the polarity of this pulse in the presence of additive white Gaussian noise (AWGN) with zero mean and $\sigma_\mathrm{n}^2$ variance can be expressed, using the complementary error function, as ${\half\erfc\left(\frac{-|A|}{\sigma_\mathrm{n}\sqrt{2}}\right)}$. Then the pulses with the magnitude ${|A|>\sigma_\mathrm{n}\sqrt{2}\erfc^{-1}(2\varepsilon)}$ will have a pulse identification error rate smaller than~$\varepsilon$. For example, $\varepsilon\lesssim 1.3\!\times\! 10^{-3}$ for $|A|\gtrsim 3\sigma_\mathrm{n}$, and $\varepsilon\lesssim 3.2\!\times\! 10^{-5}$ for $|A|\gtrsim 4\sigma_\mathrm{n}$.

The pulse rate in a digitally sampled train with regular (periodic) arrival times is $\mathcal{R}=F_\mathrm{s}/N_\mathrm{p}$, where $F_\mathrm{s}$ is the sampling frequency and $N_\mathrm{p}$ is the number of samples between two adjacent pulses in the train. For~${\mathcal{R}}$ that is sufficiently smaller than ${\mathcal{R}_0}$, the PAPR of a train of equal-magnitude pulses with regular arrival times is an {\it increasing\/} function of the number of samples between two adjacent pulses~${N_\mathrm{p}}$, and is proportional to~${N_\mathrm{p}}$:
\begin{equation} \label{eq:PAPR vs rate}
  \mathrm{PAPR} = \mathrm{PAPR}(N_\mathrm{p}) \propto {N_\mathrm{p}} \quad \mbox{for large} \quad N_\mathrm{p}\,.
\end{equation}
For example, for raised-cosine (RC) pulses ${\mathcal{R}}_0\approx (4T_\mathrm{s})^{-1}$, where $T_\mathrm{s}$ is the symbol-period, and a ``large $N_\mathrm{p}$" would mean ${N_\mathrm{p}\gg T_\mathrm{s}F_\mathrm{s} = N_\mathrm{s}}$, where $N_\mathrm{s}$ is the number of samples per symbol-period. As illustrated in Fig.~\ref{fig:RC PAPR}, ${\mathrm{PAPR}(N_\mathrm{p}) \approx 1.143\, N_\mathrm{p}/N_\mathrm{s}}$ for~$N_\mathrm{p}/N_\mathrm{s}\gg 1$ for RC pulses with roll-off factor~$\beta\!=\!1/2$ and integer values of~$N_\mathrm{s}$.

\addtocounter{figure}{1}
\begin{figure}[!b]
\centering{\includegraphics[width=8.6cm]{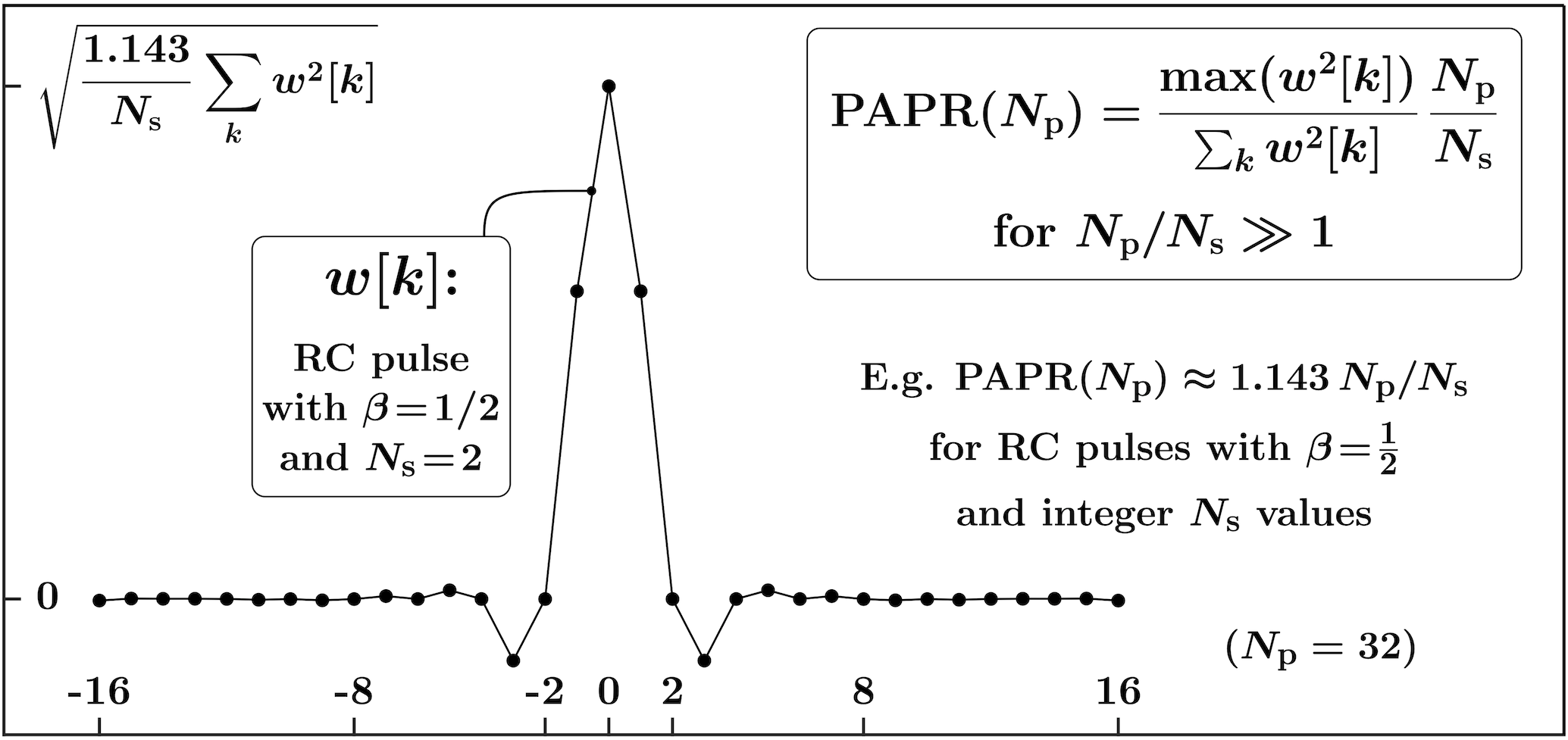}}
\caption{\boldmath PAPR for train of equal-magnitude RC pulses separated by $N_\mathrm{p} \gg N_\mathrm{s}$ is equal to PAPR of single pulse calculated on interval ${[-N_\mathrm{p}/2,N_\mathrm{p}/2]}$.
\label{fig:RC PAPR}}
\end{figure}
\begin{figure}[!t]
\centering{\includegraphics[width=8.6cm]{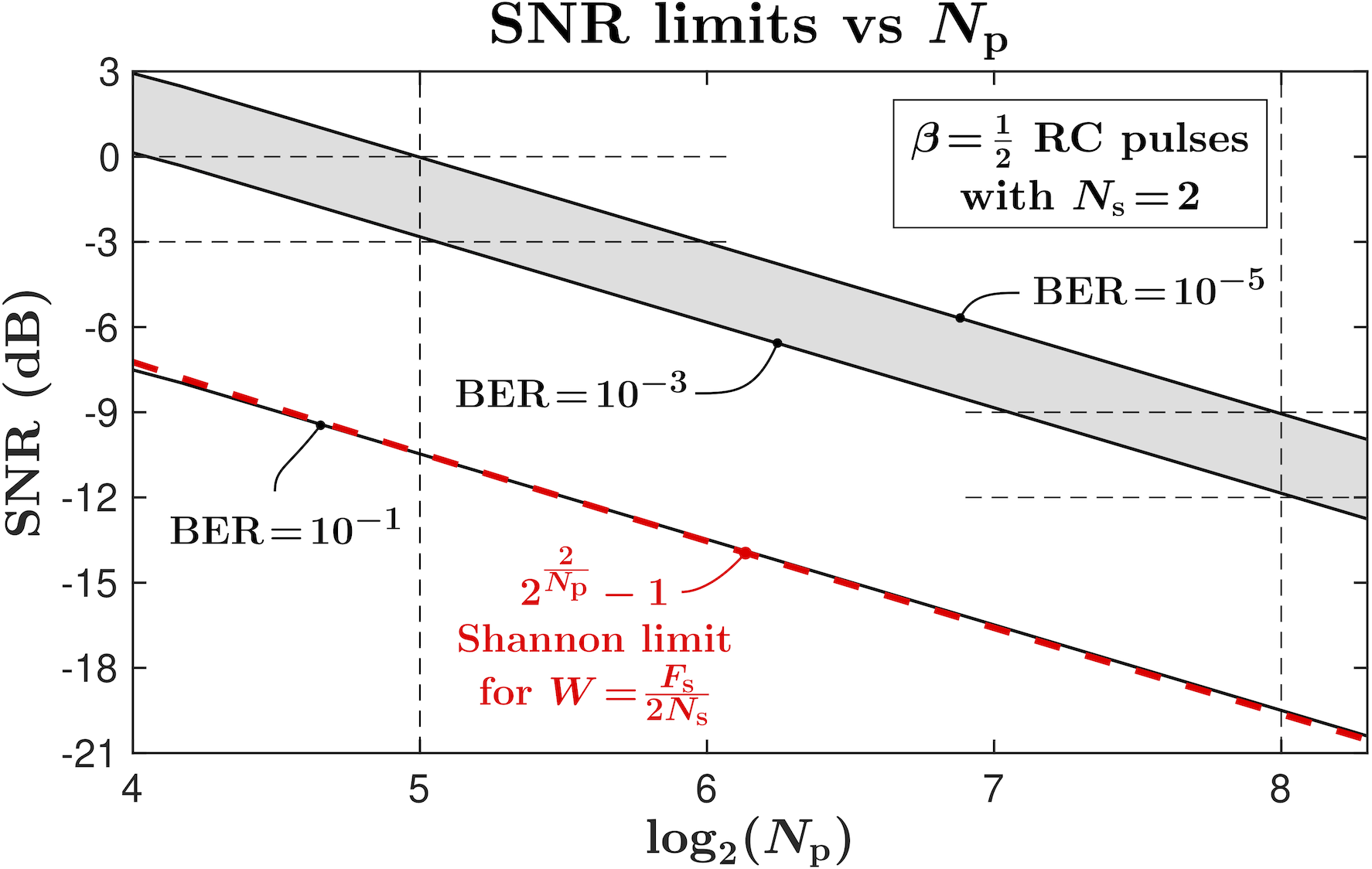}}
\caption{AWGN SNR limits for different BER as functions of samples between pulses for raised-cosine pulses with $\beta\!=\!1/2$ and $N_\mathrm{s}\!=\!2$.
\label{fig:SNR limits}}
\end{figure}
\begin{figure*}[!b]
\centering{\includegraphics[width=12.4cm]{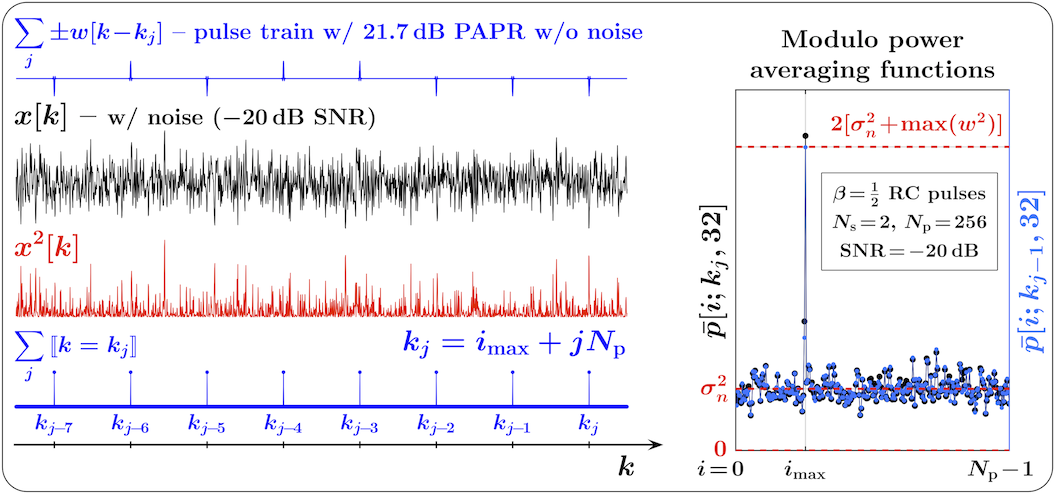}}
\caption{Illustration of synchronization procedure described by (\ref{eq:p bar}--\ref{eq:i max}). AWGN ${\mathrm{SNR}}=-20\,$dB is chosen to be low, and $M\!=\!32$ respectively high, to emphasize robustness even when ${\mathrm{BER}}\approx 1/3$.
\label{fig:synchronization}}
\end{figure*}

From the lower limit on the magnitude of a pulse for a given uncoded bit error rate (BER),
\begin{equation} \label{eq:A synch}
  |A| = \sigma_\mathrm{n} \sqrt{\mathrm{SNR}\!\times\! \mathrm{PAPR}} > \sigma_\mathrm{n}\sqrt{2}\erfc^{-1}(2\times\mathrm{BER})\,,
\end{equation}
we can obtain the lower limit on the SNR for a given pulse rate:
\begin{equation} \label{eq:SNR vs Np}
  \mathrm{SNR}(N_\mathrm{p};\mathrm{BER}) > \frac{2\left[ \erfc^{-1}(2\times\mathrm{BER}) \right]^2}{\mathrm{PAPR}(N_\mathrm{p})} \propto {N_\mathrm{p}^{-1}},
\end{equation}
or
\begin{equation} \label{eq:SNR vs Np RC}
  \mathrm{SNR}(N_\mathrm{p};\mathrm{BER}) \gtrsim 1.75\left[ \erfc^{-1}(2\times\mathrm{BER}) \right]^2 \frac{N_\mathrm{s}}{N_\mathrm{p}}
\end{equation}
for~$N_\mathrm{s}/N_\mathrm{p}\ll 1$ and RC pulses with~$\beta\!=\!1/2$. For example, $\mathrm{SNR}(N_\mathrm{p};10^{-3}) \gtrsim 9.6/\mathrm{PAPR}(N_\mathrm{p}) \approx 8.4\, N_\mathrm{s}/N_\mathrm{p}$, and $\mathrm{SNR}(N_\mathrm{p};10^{-5}) \gtrsim 18.2/\mathrm{PAPR}(N_\mathrm{p})  \approx 15.9\, N_\mathrm{s}/N_\mathrm{p}$.

Fig.~\ref{fig:SNR limits} illustrates the SNR limits for different BER as functions of samples between pulses for RC pulses with $\beta\!=\!1/2$ and $N_\mathrm{s}\!=\!2$. For example, for the pulses separated by 128~symbol-periods, $\mathrm{BER}
\!\lesssim\!10^{-3}$ is achieved for $\mathrm{SNR}\!\gtrsim\!-12\,$dB. For comparison, the AWGN Shannon capacity limit~\cite{Shannon49communication} for the bandwidth~$W\!=\!F_\mathrm{s}/(2N_\mathrm{s})$, which is the nominal bandwidth of the respective RRC filter, is also shown.

\subsection{Asynchronous Detection (Pulse Counting)} \label{subsec:counting}
When the time intervals between the pulses in a train do not compose a periodic structure, one must use asynchronous detection (pulse counting). In pulse counting, a pulse is detected when it crosses a certain non-zero threshold. A {\it false positive\/} (fp) detection occurs when such crossing is entirely due to noise, and a {\it false negative\/} (fn) detection happens when a pulse affected by the noise fails to cross the threshold. For a positive threshold~$\alpha^+>0$, the false negative rate will be smaller than some tolerance rate~$\varepsilon_{\mathrm{fn}}$ if the amplitude of a pulse is ${A>\alpha^+ + \sigma_{\mathrm{n}}\sqrt{2}\erfc^{-1}(2\varepsilon_{\mathrm{fn}})}$.

As shown in~\cite{Rice44and45mathematical, Nikitin98a}, for a filtered noise with zero mean and $\sigma_{\mathrm{n}}^2$~variance, its rate of up-crossing the threshold~$\alpha^+>0$ can be expressed as ${{\mathcal{R}}_{\mathrm{max}}\exp \left( -\half(\alpha^+/\sigma_{\mathrm{n}})^2\right)}$, where the {\it saturation rate\,}~${\mathcal{R}}_{\mathrm{max}}$ is determined entirely by the filter's frequency response. Then, for the average pulse arrival rate~${\mathcal{R}}$, the threshold value needs to be $\alpha^+ > \sigma_{\mathrm{n}} \left[ -2\ln(\varepsilon_{\mathrm{fp}} {\mathcal{R}}/{\mathcal{R}}_{\mathrm{max}}) \right]^\half$ in order to keep the false positive rate below~$\varepsilon_{\mathrm{fp}}$. For example, for ${\mathcal{R}}/{\mathcal{R}}_{\mathrm{max}}=1/10$, the threshold value is $\alpha^+ \gtrsim 4.3\sigma_{\mathrm{n}}$ for $\varepsilon_{\mathrm{fp}}=10^{-3}$, and $\alpha^+ \gtrsim 4.8\sigma_{\mathrm{n}}$ for $\varepsilon_{\mathrm{fp}}=10^{-4}$. Note that, as shown in~\cite{Rice44and45mathematical}, for an ideal ``brick wall" lowpass filter with the bandwidth~$\Delta{B}$ the saturation rate ${\mathcal{R}}_{\mathrm{max}}=\Delta{B}/\sqrt{3}$. Hence, for example, for a root-raised-cosine or a raised-cosine filter ${\mathcal{R}}_{\mathrm{max}}\approx (2T_{\mathrm{s}}\sqrt{3})^{-1}$, where~$T_{\mathrm{s}}$ is the reciprocal of the symbol-rate parameter of the filter. For a pulse rate~${\mathcal{R}}$ that is sufficiently smaller than ${\mathcal{R}}_0\!=\! \half\Delta{B}/{\mathrm{TBP}}$, the PAPR of a train of equal-magnitude pulses is inversely proportional to~${\mathcal{R}}$. Then, for a given signal-to-noise ratio (SNR) of a pulse train affected by additive Gaussian noise, and for a given false negative rate constraint~$\varepsilon_{\mathrm{fn}}$, the pulse rate needs to be sufficiently small to ensure the pulse detection with the error rate below~$\varepsilon_{\mathrm{fn}}$.

The asynchronous pulse detection (pulse counting) is discussed in more detail in Section~\ref{sec:INF}. While the rate limit for pulse counting is approximately an order of magnitude lower than for synchronous pulse detection with a similar BER, pulse counting does not rely on any {\it a priori\/} knowledge of pulse arrival times, and can be used as a backbone method for pulse detection. In addition, randomizing the pulse arrival times allows us to more effectively hide the temporal structure of the pulse train, prioritizing security over the data rates. Further, intermittently nonlinear filtering (INF) used in combination with synchronous and/or asynchronous pulse detection enables ``layering" of pulse trains with significantly different powers, physical-layer steganography, and ``friendly jamming" applications. However, since synchronous detection enables much higher data rates for the same SNR, the focus of the next section is on the technique that can be used for synchronous detection of pulses in a train with a periodic structure of interarrival times. In practice, both pulse counting and synchronous pulse detection can be used in combination. For example, given a constraint on the total power of the pulse train, counting of relatively rare, higher-magnitude pulses can be used to establish the timing patterns for synchronization, and synchronous detection of smaller, more frequent pulses can be used for a higher data rate.

\section{Synchronization for regular pulse trains} \label{sec:synchronization}
Let us consider the basic link shown in Fig.~\ref{fig:low SNR}, where in the transmitter a binary sequence is encoded in a periodic pulse train according to~(\ref{eq:ptrain equidistant}), with ${N=N_{\mathrm{p}}}$. To recover this sequence with minimal raw BER, we would need to sample the received pulse train $x[k]$ at the instances where the pulses of the ``ideal" (without noise) pulse train have maximum magnitudes. To determine the respective sampling indexes, the following {\it modulo power averaging\/} (MPA) function can be constructed as an exponentially decaying average of the instantaneous signal power~$x^2[k]$ in a window of size~$N_{\mathrm{p}}\!+\!1$:
\beginlabel{align}{eq:p bar}
  &\bar{\mathrm{p}}[i;k_j,M] = \frac{M\!-\!1}{M}\, \bar{\mathrm{p}}[i;k_{j\!-\!1},M]\\[1ex]
  &+ \frac{1}{M} \sum_k \Ibl k\!\ge\! k_j\!-\!N_{\mathrm{p}}\Ibr \Ibl k\!\le\! k_j\Ibr \Ibl i\!=\!\mod(k,N_{\mathrm{p}})\Ibr\, x^2[k]\,,\nonumber
\end{align}
where $k_j$ is the sample index of the $j$-th pulse, and~${M\!>\!1}$. Thus the window $k_j\!-\!N_{\mathrm{p}}\!\le\!k\!\le\! k_j$ includes two transmitted pulses, $k_{j\!-\!1}$ and $k_j$, and the index~$i$ in $\bar{\mathrm{p}}[i;k_j,M]$ takes the values $i=0,\ldots,N_{\mathrm{p}}\!-\!1$. Note that using exponentially decaying average in~(\ref{eq:p bar}) would roughly correspond to averaging $N\!\approx\!2M\!-\!1$ such windows. The exponentially decaying average, however, has the advantage of lower computational and memory burden, especially for large~$M$, and faster adaptability to dynamically changing conditions. 

For a sufficiently large~$M$, the peak in $\bar{\mathrm{p}}[i;k_j,M]$ corresponding to the pulses of the ideal pulse train will dominate. Therefore, the index~$k_{j\!+\!1}$ for sampling of the $(j\!+\!1)$-th pulse can be obtained as
\beginlabel{equation}{eq:kj}
  k_{j\!+\!1} = i_{\mathrm{max}} + ({j\!+\!1}) N_{\mathrm{p}}\,,
\end{equation}
where $i_{\mathrm{max}}$ is given by
\beginlabel{equation}{eq:i max}
  \bar{\mathrm{p}}[i_{\mathrm{max}};k_j,M] = \max_i\left( \bar{\mathrm{p}}[i;k_j,M] \right).
\end{equation}
Note that reliance on the global maximum of~$\bar{\mathrm{p}}[i;k_j,M]$ is also likely to provide resilience to multipath interference, as this maximum will be mainly determined by the dominating (e.g. a line-of-sight) signal component.

Fig.~\ref{fig:synchronization} illustrates this synchronization procedure. The MPA function shown on the right-hand side of the figure is computed according to~(\ref{eq:p bar}). To emphasize the robustness of this synchronization technique even when the bit error rates are very high, the SNR is chosen to be respectively low (${\mathrm{SNR}}=-20\,$dB, ${\mathrm{BER}}\approx 1/3$ in this example).

\begin{figure}[!t]
\centering{\includegraphics[width=8.6cm]{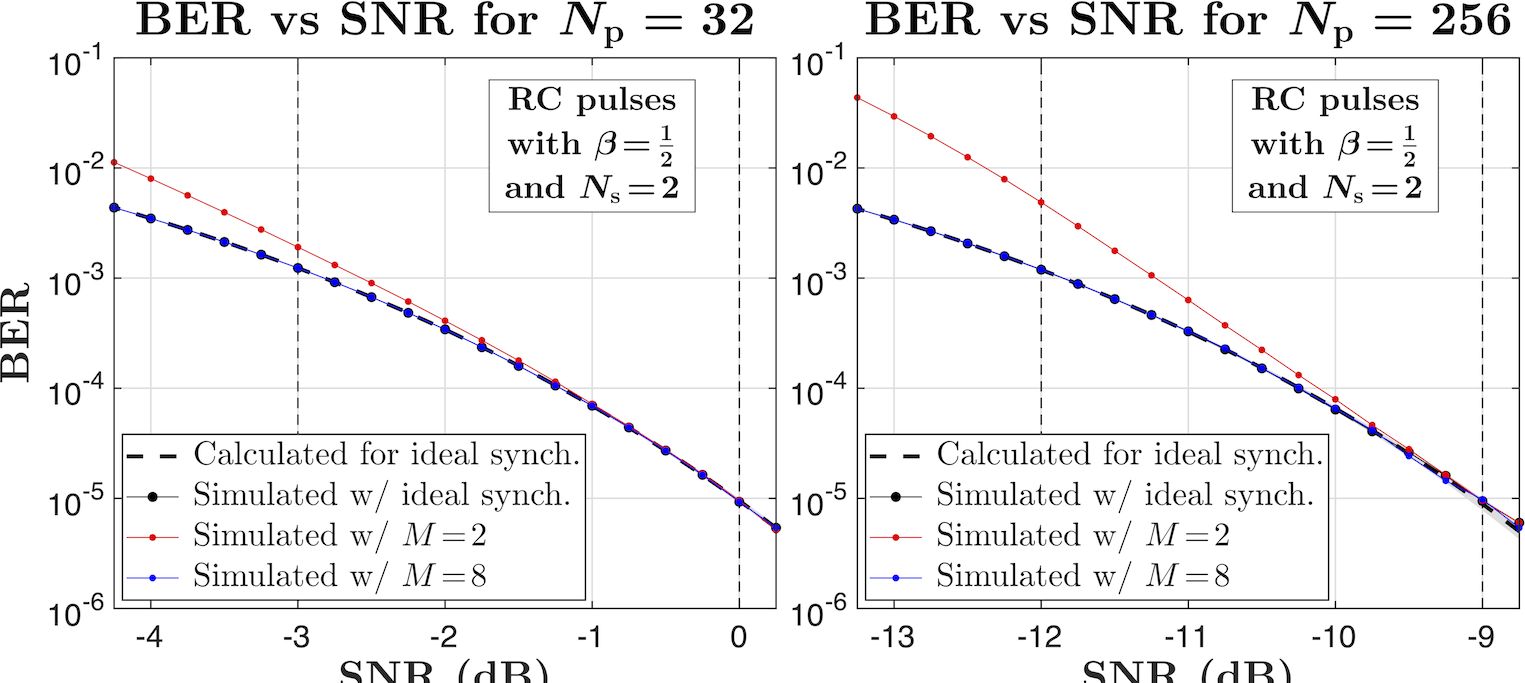}}
\caption{Calculated and simulated BERs as functions of AWGN SNRs for $N_{\mathrm{p}}=32$ and $N_{\mathrm{p}}=256$. For shown SNR ranges, MPA function with ${M\!=\!8}$ provides reliable synchronization. (Compare with SNR limits in Fig.~\ref{fig:SNR limits}.)
\label{fig:BER vs SNR}}
\end{figure}

For the link shown in Fig.~\ref{fig:low SNR}, and for the RC pulses with $\beta\!=\!1/2$ and $N_\mathrm{s}\!=\!2$ used in Section~\ref{sec:pulsed waveforms}, Fig.~\ref{fig:BER vs SNR} compares the calculated (dashed lines) and the simulated (dots connected by solid lines) BERs, for the ``ideal" synchronization (black dots), and for the synchronization with the MPA function described above. The AWGN noise is added at the receiver input, and the SNR is calculated at the output of the matched filter in the receiver. One can see that for ${M\!=\!2}$ (red dots) the errors in synchronization are relatively high, which increases the overall BER, but the MPA function with ${M\!=\!8}$  (blue dots) provides reliable yet still fast synchronization. The BERs and the respective SNRs in Fig.~\ref{fig:BER vs SNR} are presented for the pulse repetition rates indicated by the vertical dashed lines in Fig.~\ref{fig:SNR limits}.

\begin{figure}[!b]
\centering{\includegraphics[width=8.6cm]{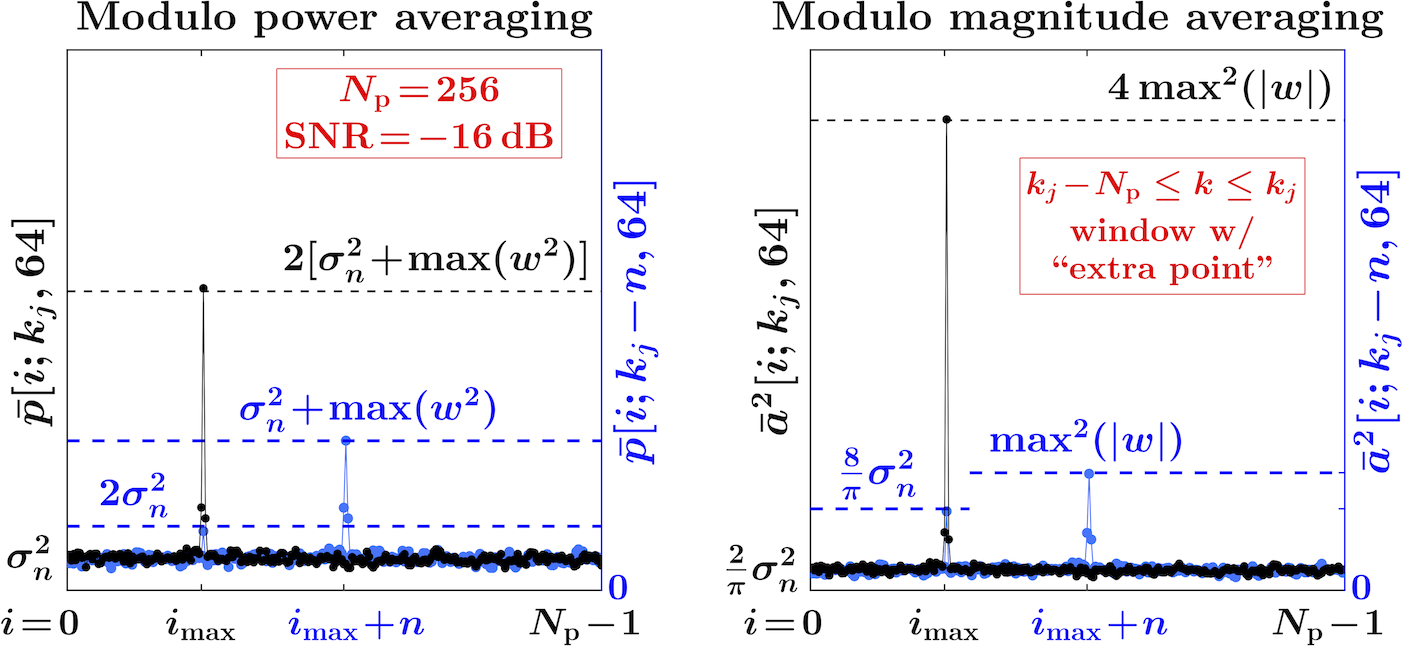}}
\caption{If used in modulo magnitude averaging, ``extra point" significantly increases probability of synchronization failure.
\label{fig:synchronization abs}}
\end{figure}

\subsection{Modulo magnitude averaging} \label{subsec:MMA}
When a pulse train is used for communications rather than, say, radar applications, reliable synchronization may only need to be achievable for relatively low BER, e.g. ${{\mathrm{BER}}\lesssim 1/10}$. Then the following {\em modulo magnitude averaging\/} (MMA) function can replace the MPA function in the synchronization procedure, in order to reduce the computational burden by avoiding squaring operations:
\beginlabel{align}{eq:a bar}
  &\bar{\mathrm{a}}[i;k_j,M] = \frac{M\!-\!1}{M}\, \bar{\mathrm{a}}[i;k_{j\!-\!1},M]\\[1ex]
  &+ \frac{1}{M} \sum_k \Ibl k\!>\! k_j\!-\!N_{\mathrm{p}}\Ibr \Ibl k\!\le\! k_j\Ibr \Ibl i\!=\!\mod(k,N_{\mathrm{p}})\Ibr\, \left|x[k]\right|\,.\nonumber
\end{align}
Note that the window $k_j\!-\!N_{\mathrm{p}}\!<\!k\!\le\! k_j$ in~(\ref{eq:a bar}) includes only the $j$-th transmitted pulse, instead of two pulses used in~(\ref{eq:p bar}). The reason behind this is illustrated in Fig.~\ref{fig:synchronization abs}, which compares (for AWGN) the MPA function $\bar{\mathrm{p}}[i;k_j,M]$ with the respective squared MMA function $\bar{\mathrm{a}}^2[i;k_j,M]$ computed for the window $k_{j}\!-\!N_{\mathrm{p}}\!\le\!k\!\le\! k_{j}$ that includes the ``extra point" (the $(j\!-\!1)$-th pulse). The relatively long averaging ($M\!=\!64$) is used to reduce the variations in the function values due to noise, and to make the comparison with the levels indicated by the dashed lines more apparent.

\begin{figure}[!b]
\centering{\includegraphics[width=8.6cm]{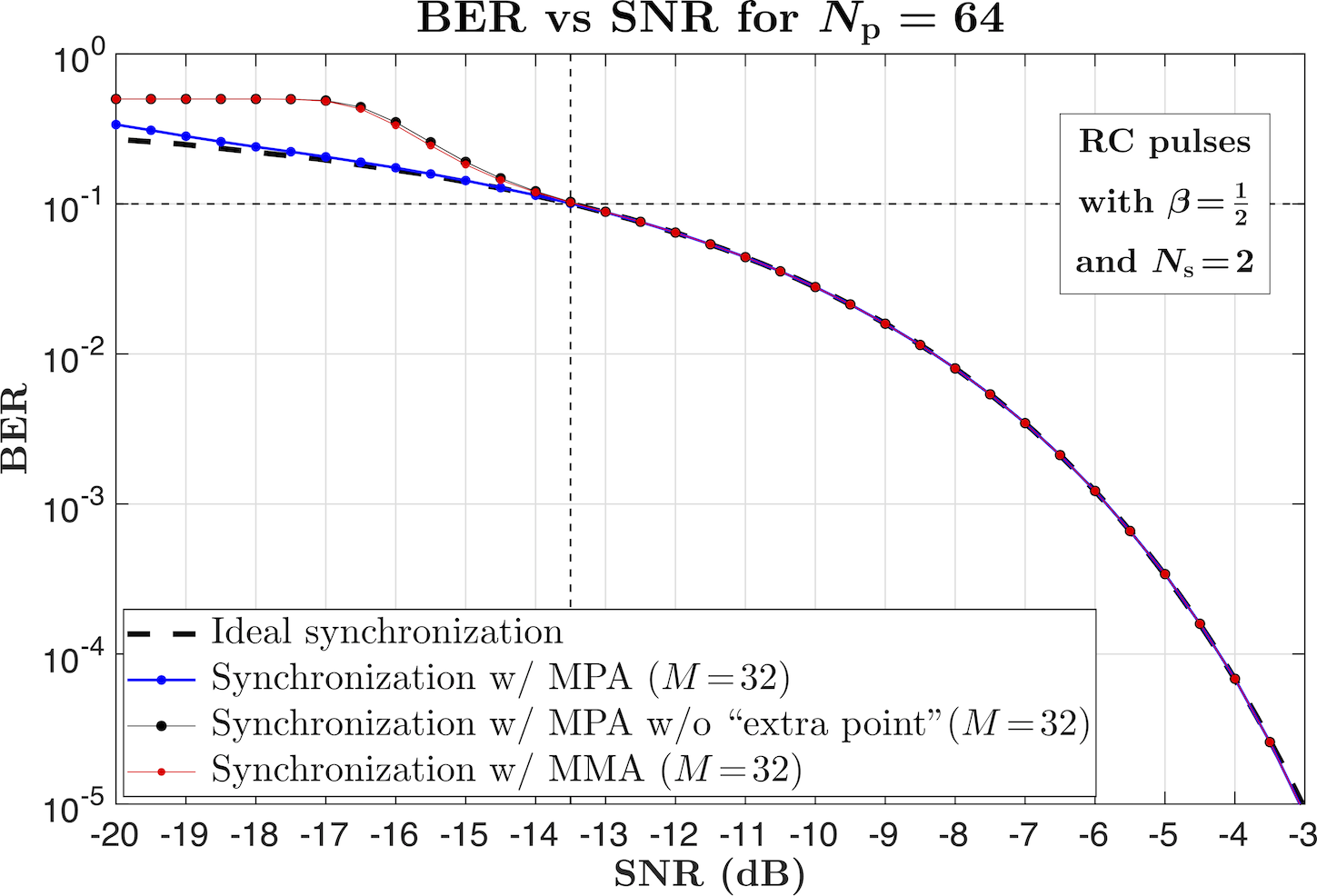}}
\caption{For BER smaller than about $10^{-1}$, less computationally expensive modulo magnitude averaging (e.g. given by~(\ref{eq:a bar})) can be used for synchronization. Modulo power averaging (with ``extra point," e.g. given by~(\ref{eq:p bar})) should be used when reliable synchronization for full BER range is desired.
\label{fig:synchronization comparison}}
\end{figure}

When a correct synchronization has already been obtained, and the maxima are ``locked" at the correct $i_{\mathrm{max}}$ values (black dots connected by solid lines in Fig.~\ref{fig:synchronization abs}), both the MPA and the MMA functions would adequately maintain the position of their maxima. However, an offset in the synchronization (e.g. by $n$ points shown in the figure) significantly more unfavorably affects the margin between the extrema at $i_{\mathrm{max}}$ and $i_{\mathrm{max}}\!+\!n$ in the MMA function, compared with the MPA function (blue dots connected by solid lines). Thus the ``extra point" may cause the ``failure to synchronize" even at a relatively high SNR, and it should be removed from the calculation of the MMA function. Then, as illustrated in Fig.~\ref{fig:synchronization comparison}, for ${{\mathrm{BER}}\lesssim 1/10}$ synchronization with the MMA function $\bar{\mathrm{a}}[i;k_j,M]$ would be effectively equivalent to synchronization with the MPA function $\bar{\mathrm{p}}[i;k_j,M]$. When reliable synchronization for larger BERs is desired (e.g. in timing and ranging applications), then the MPA  given by~(\ref{eq:p bar}) should be used.

\section{Intermittently Nonlinear Filtering (INF) for Outlier Noise Mitigation and Pulse Counting} \label{sec:INF}
In addition to ever-present thermal noise (which can be appropriately modeled as AWGN), the received signal can contain significant amounts of non-Gaussian interference originating from a multitude of natural and technogenic (man-made) phenomena. Then the overall noise would be non-Gaussian and, depending on the noise coupling mechanisms and the system's filtering properties and propagation conditions, it may contain distinct amplitude outliers when observed in the time domain. The presence of different types of such outlier noise is widely recognized in multiple applications under various general and application-specific names, most commonly as {\it impulsive\/}, {\it transient\/}, {\it burst\/}, or {\it crackling\/} noise. The outlier noise can be efficiently mitigated in real-time using intermittently nonlinear filters (INF)~\cite[e.g.]{Nikitin19hidden}, and, depending on the noise nature and composition, improvements in the quality of the signal of interest can vary from “no harm" to substantial.

Since in the power-limited regime the channel capacity is proportional to the SNR, even relatively small increase in the latter can be beneficial. For example, as can be seen in Fig.~\ref{fig:SNR limits}, for a given raw BER a $3\,$dB increase in the SNR enables doubling of the bit rate. Alternatively, for a given pulse rate, a $3\,$dB increase in the SNR can reduce the raw BER by several orders of magnitude. This is why, as illustrated in Fig.~\ref{fig:low SNR}, it would be useful to deploy INF for mitigation of outlier noise in practical implementations of low-SNR links described in this paper. Note that INF should be performed before the large-TBP filtering in the receiver, when the signal of interest is still Gaussian or sub-Gaussian, and the outlier structure of the noise is still apparent and not affected by the pileup.

\begin{figure}[!b]
\centering{\includegraphics[width=8.6cm]{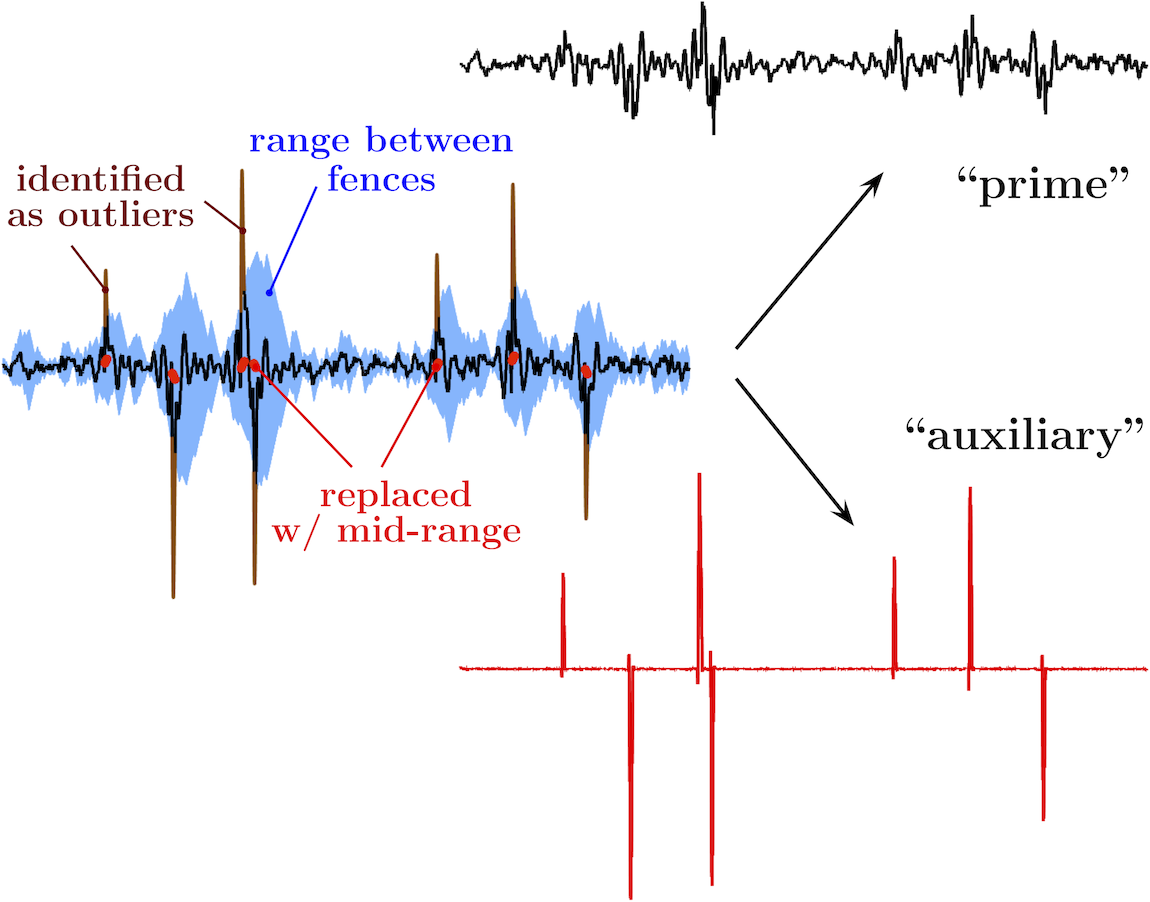}}
\caption{{\bf Intermittently Nonlinear Filtering (INF):} Outliers are identified as protrusions outside of fenced range, and their values are replaced by those in mid-range. Otherwise, signal is not affected. ``Auxiliary" output is difference between input and ``prime" INF output.
\label{fig:INF}}
\end{figure}

In general, a nonlinear filter is capable of disproportionately affecting spectral densities of signals with distinct temporal and/or amplitude structures even when these signals have the same spectral content. In particular, the separation of a large-PAPR pulse train and a small-PAPR signal can be viewed as either (i)~mitigation of impulsive noise affecting the small-PAPR signal, or (ii)~extraction of impulsive signal from the small-PAPR background. In this paper, a specific type of Intermittently Nonlinear Filters (INF) is used to accomplish either or both tasks. While various INF configurations, their different uses, and the approaches to their analog and/or digital implementations are described elsewhere~\cite{Nikitin19ADiCpatentCIP2, Nikitin19complementary, Nikitin19hidden, Nikitin19quantile, Nikitin18ADiC-ICC}, Fig.~\ref{fig:INF} illustrates their basic concept. In an INF, the upper and the lower fences establish a robust range that excludes high-amplitude pulses while effectively containing the small-PAPR component. The prime INF output simply contains the input signal in which the outliers (i.e. the pulses that protrude from the range) are replaced with mid-range values. This constitutes mitigation of impulsive noise affecting the small-PAPR signal. The auxiliary INF output is the difference between its input and the prime output. This is akin to extraction of impulsive signal from the small-PAPR background (or ``pulse counting").

\begin{figure}[!b]
\centering{\includegraphics[width=8.6cm]{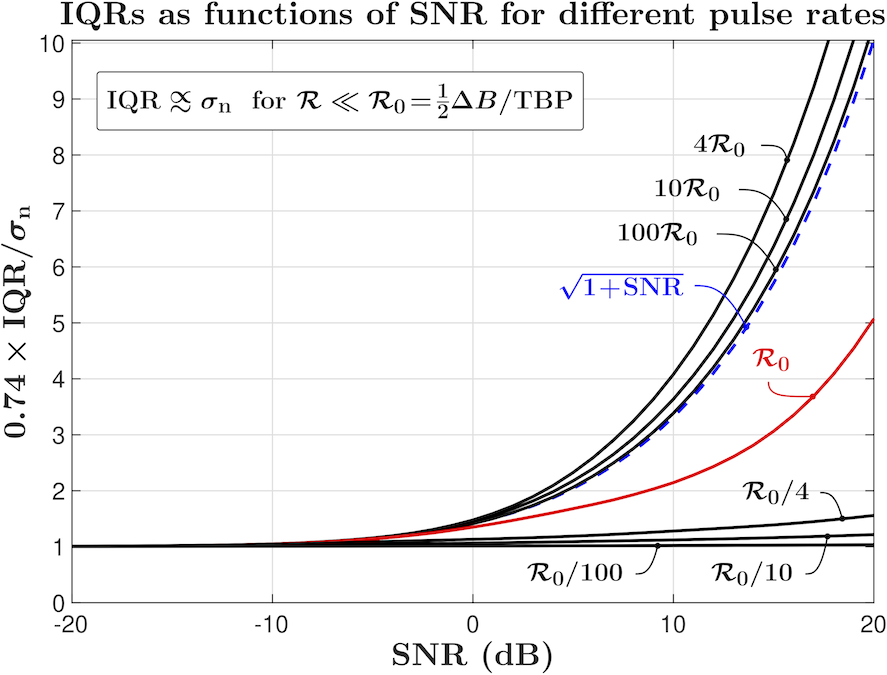}}
\caption{For low pulse rates (e.g. ${\mathcal{R}}\!\ll\! \half\Delta{B}/{\mathrm{TBP}}$), IQR provides reliable measure of additive Gaussian noise power, $\sigma_{\mathrm{n}}\propto {\mathrm{IQR}}$. Root-raised-cosine pulses (for which ${\mathcal{R}}_0\approx (4{T_{\mathrm{s}}})^{-1}$) are used in this example. For completeness, IQRs for higher rates are also shown, but details of their change with SNR are not discussed.
\label{fig:IQR}}
\end{figure}

\subsection{Robust Range/Fencing in INF} \label{subsec:IQR}
For an INF to be effective in separation of small-PAPR and impulsive signals regardless of their relative powers, its range needs to be robust (insensitive) to the pulse train. Favorably, for a mixture of a small-PAPR signal with the bandwidth~$\Delta{B}$, and a pulse train with the same bandwidth and the rate sufficiently below ${\mathcal{R}}_0$, when pileup effect is insignificant, the value of the interquartile range (IQR) of the mixture is insensitive to the power of the pulse train. This is illustrated in Fig.~\ref{fig:IQR} for a pulse train affected by additive Gaussian noise. Thus robust upper ($\alpha^+$) and lower ($\alpha^-$) fences for INF can be constructed as linear combinations of the 1st ($Q_{[1]}$) and the 3rd ($Q_{[3]}$) quartiles of the signal (Tukey's fences~\cite{Tukey77exploratory}) obtained in a moving time window:
\begin{equation} \label{eq:Tukey's range}
  [\alpha^-,\alpha^+] = {\left[Q_{[1]}\!-\!\beta\left(Q_{[3]}\!-\!Q_{[1]}\right)\!,\,Q_{[3]}\!+\!\beta\left(Q_{[3]}\!-\!Q_{[1]}\right)\!\right]},
\end{equation}
where $\alpha^+$, $\alpha^-$, $Q_{[1]}$, and $Q_{[3]}$ are time-varying quantities, and $\beta$ is a scaling parameter of order unity. When an INF is used for pulse counting in the presence of AWGN, the particular value of~$\beta$ may be chosen based on the constraint on the relative rate~$\varepsilon_{\mathrm{fp}}$ of false positive detections. Then, as follows from the discussion in Section~\ref{subsec:counting},
\begin{equation} \label{eq:beta}
\beta \approx 1.05\times \sqrt{ \ln\left( \frac{{\mathcal{R}}_{\mathrm{max}}}{\varepsilon_{\mathrm{fp}} {\mathcal{R}}} \right)} \,- \half\,.
\end{equation}
For example, for ${\mathcal{R}}_{\mathrm{max}}/{\mathcal{R}}=10$, $\beta \approx 2.7$ for $\varepsilon_{\mathrm{fp}}=10^{-3}$, and $\beta \approx 3.1$ for $\varepsilon_{\mathrm{fp}}=10^{-4}$.

\begin{figure}[!t]
\centering{\includegraphics[width=8.6cm]{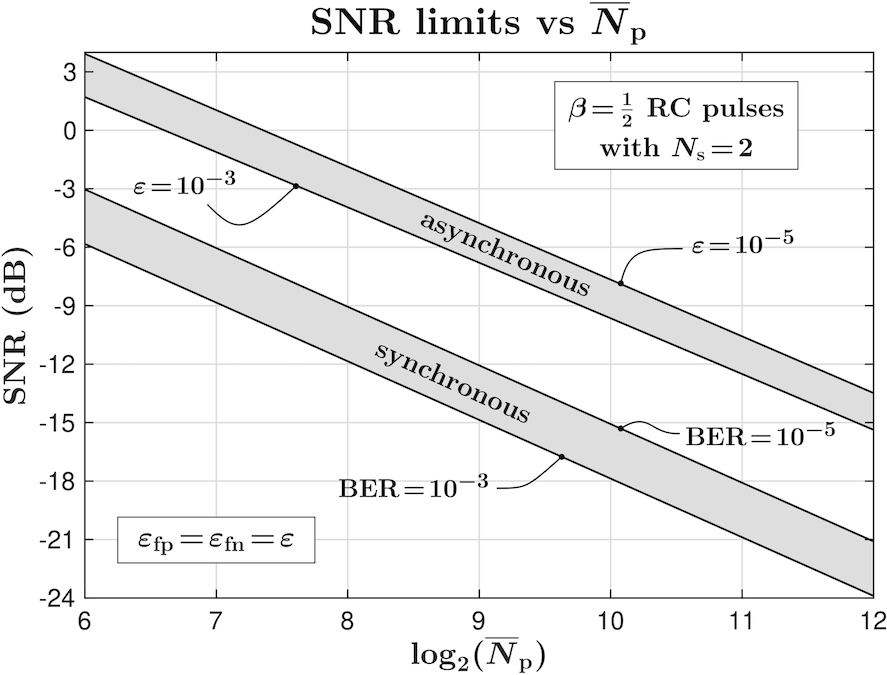}}
\caption{Illustrative comparison of SNR limits under AWGN for synchronous and asynchronous detection. Trains of equal-magnitude RC pulses with roll-off factor~$1/2$ and with average interarrival intervals~$\overline{N}_{\mathrm{p}} \gg N_\mathrm{s}$ are used.
\label{fig:SNR limits async}}
\end{figure}

For Tukey's fences with $\beta$ given by~(\ref{eq:beta}), the probability of a false negative count of a pulse with the magnitude~$|A|$ can be expressed as
\begin{equation} \label{eq:fn async}
  \varepsilon_{\mathrm{fn}} \approx \half\erfc\left(
    \frac{|A|}{\sigma_{\mathrm{n}}\sqrt{2}} -
    \sqrt{ \ln\left( \frac{{\mathcal{R}}_{\mathrm{max}}}{\varepsilon_{\mathrm{fp}} \mathcal{R}}\right) }
  \right).
\end{equation}
Then, for example, for a train of equal-magnitude RC pulses with the roll-off factor~$1/2$, and with the average interarrival interval~$\overline{N}_{\mathrm{p}} \gg N_\mathrm{s}$ (with a possible constraint on the minimal interarrival distance to avoid pileup), the lower limit on the SNR for a given average pulse rate $1/\overline{N}_{\mathrm{p}}$ can be expressed as
\begin{align} \label{eq:SNR vs Np RC async}
  \mathrm{SNR}&(\overline{N}_\mathrm{p};\varepsilon_{\mathrm{fp}},\varepsilon_{\mathrm{fn}}) \gtrsim \nonumber\\
  &1.75\left[ \erfc^{-1}(2\varepsilon_{\mathrm{fn}}) + \sqrt{ \ln\left( \frac{\overline{N}_\mathrm{p}}{3.5\, \varepsilon_{\mathrm{fp}} N_\mathrm{s}} \right)} \right]^2 \!\frac{N_\mathrm{s}}{\overline{N}_\mathrm{p}}\,.
\end{align}

Fig.~\ref{fig:SNR limits async} provides illustrative comparison of the AWGN SNR limits for synchronous and asynchronous detection. Note that while the rate limit for asynchronous detection is almost an order of magnitude lower than that for synchronous with a similar error rate, randomizing the pulse arrival times allows us to more effectively hide the temporal structure of the pulse train by a large-TBP pulse shaping, prioritizing security over the data rates.

\subsection{Quantile Tracking Filters for Robust Fencing} \label{subsec:QTFs}
As a practical matter, quantile tracking filters (QTFs) are an appealing choice for such robust fencing in INF, as QTFs are analog filters suitable for wideband real-time processing of continuous-time signals and are easily implemented in analog circuitry~\cite{Nikitin19ADiCpatentCIP2, Nikitin19complementary, Nikitin19hidden, Nikitin19quantile, Nikitin18ADiC-ICC}. Further, their numerical computations are $\mathcal{O}(1)$ per output value in both time and storage, which also enables their high-rate digital implementations in real time.

\begin{figure}[!b]
\centering{\includegraphics[width=8.6cm]{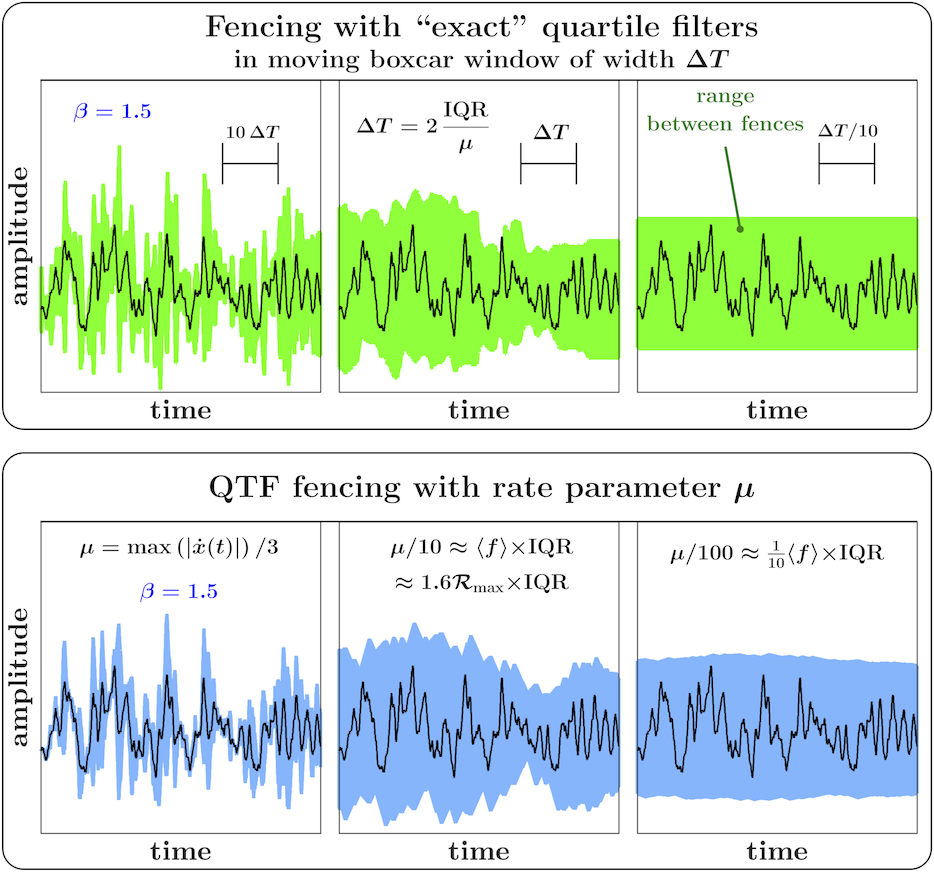}}
\caption{Overall behavior of QTF fencing is similar to that with ``exact" quartile filters in moving boxcar window of width~$\Delta{T}=2\times{\mathrm{IQR}}/\mu$.
\label{fig:RANKvsQTFs}}
\end{figure}

In brief, the signal~$Q_q(t)$ that is related to the given input~$x(t)$ by the equation
\begin{equation} \label{eq:QTF eps}
  \frac{\d}{\d{t}}\, Q_q = \mu\, \left[\lim_{\varepsilon\to 0}\Seps(x\!-\!Q_q) + 2q-1\right],
\end{equation}
where $\mu$ is the {\it rate parameter\/} and $0\!<\!q\!<\!1$ is the {\it quantile parameter\/}, can be used to approximate (``track") the $q$-th~quantile of $x(t)$ for the purpose of establishing a robust range ${[\alpha^-,\alpha^+]}$. In~(\ref{eq:QTF eps}), the {\it comparator function\,}~$\Seps(x)$ can be any {\it continuous\/} function such that~$\Seps(x)=\sgn(x)$ for~$|x|\gg\varepsilon$, and $\Seps(x)$~changes monotonically from~``$-1$" to~``$1$" so that most of this change occurs over the range~$[-\varepsilon,\varepsilon]$. As discussed in detail in~\cite{Nikitin19quantile}, for a continuous stationary signal~$x(t)$ with a constant mean and a positive IQR, the outputs~$Q_{[1]}(t)$ and~$Q_{[3]}(t)$ of QTFs with a sufficiently small rate parameter~$\mu$ will approximate the~1st and the~3rd quartiles, respectively, of the signal obtained in a moving boxcar time window with the width~$\Delta{T}$ of order~${2\times{\mathrm{IQR}}/\mu\gg\langle f\rangle^{-1}}$, where~$\langle f\rangle$ is the average crossing rate of~$x(t)$ with the~1st and the~3rd quartiles of~$x(t)$. Consequently, as illustrated in Fig.~\ref{fig:RANKvsQTFs}, the overall behavior of the QTF fencing for a stationary constant-mean signal with a given IQR would be similar to the fencing with the ``exact" quartile filters in a moving boxcar window $\left[ \theta(t) - \theta(t\!-\!\Delta{T}) \right]/\Delta{T}$, where~${\Delta{T}=2\times{\mathrm{IQR}}/\mu}$ and~$\mu$ is the QTF rate parameter. However, for a sampling rate~$F_{\mathrm{s}}$, numerical computations of an ``exact" quartile require $\mathcal{O}\left(F_{\mathrm{s}}\Delta{T}\log(F_{\mathrm{s}}\Delta{T})\right)$ per output value in time, and $\mathcal{O}(F_{\mathrm{s}}\Delta{T})$ in storage, becoming prohibitively expensive for high-rate real-time processing.

\subsection{Asynchronous Sampling of Pulse Trains} \label{subsec:asynchronous sampling}
Let us assume that the designed pulse train~$\hat{x}[k]$ in the transmitter can be represented by~(\ref{eq:ptrain}), so that only $k_j$-th samples have non-zero values~$A_j$, and that the information is encoded by the amplitudes~$A_j$ and/or the interarrival times ${k_j\!-\!k_{j\!-\!1}}$. Further, the impulse response~$w[k]$ of the  small-TBP filter has a dominant peak, and the pulse arrival rate is small so that pileup is negligible. When a simple binary sequence is transmitted (say, ``0" or ``1" for negative or positive pulses), it can be recovered by measuring, for appropriately chosen thresholds $\alpha^+\!>\!0$ and $\alpha^-\!<\!0$, the {\it upward\/} and the {\it downward\/} crossings, respectively, of~$\alpha^+$ and~$\alpha^-$ by the pulse train~$x[k]$.

\begin{figure}[!b]
\centering{\includegraphics[width=8.6cm]{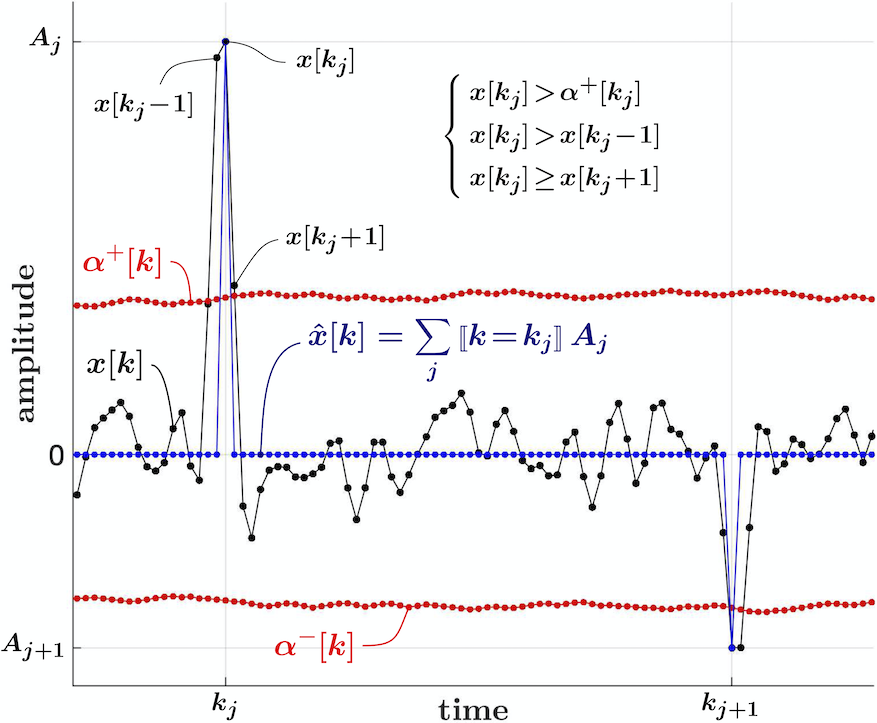}}
\caption{Sampling waveform~$x[k]$ at peaks of pulses protruding from range formed by Tukey's fences~${\left[ \alpha^-_k,\alpha^+_k \right]}$. Pulse train $\hat{x}[k]$ is computed according to~(\ref{eq:ccount sampling}).
\label{fig:pulse counting}}
\end{figure}

The threshold crossings, however, do not characterize the amplitudes of the pulses. Also, the front edges of pulses with different magnitudes would have different slew rates, thus contributing to timing arrows when the level crossings are used for timing. Instead, to extract the information about the amplitudes~$A_j$ in the train, and to reduce the error in obtaining the interarrival times ${k_j\!-\!k_{j\!-\!1}}$, the following {\it pulse counting} function~$\CalphaPM\! \left(x[k]\right)$ can be used:
\beginlabel{align}{eq:ccount}
  \CalphaPM\! \left(x_k\right) &= \Ibl x_k\!>\!\alpha^+_k\Ibr \Ibl x_k\!>\!x_{k\!-\!1}\Ibr \Ibl x_k\!\ge\!x_{k\!+\!1}\Ibr \nonumber\\
  &\,+ \Ibl x_k\!<\!\alpha^-_k\Ibr \Ibl x_k\!<\!x_{k\!-\!1}\Ibr \Ibl x_k\!\le\!x_{k\!+\!1}\Ibr\,.
\end{align}
This function takes unit values at the local extrema of $x[k]$ that protrude from the range ${\left[ \alpha^-_k,\alpha^+_k \right]}$, and is zero otherwise. (In~(\ref{eq:ccount}), for better readability, we use $x_k$ and $\alpha^{\pm}_k$ in place of $x[k]$ and $\alpha^{\pm}[k]$.) Then~$\hat{x}[k]$, obtained as a product of the pulse train~$x[k]$ and the respective counting function,
\beginlabel{equation}{eq:ccount sampling}
  \hat{x}[k] = \CalphaPM\! \left(x[k]\right) \,x[k] = \sum_j \Ibl k\!=\!k_j\Ibr\, A_j\,,
\end{equation}
will represent sampling of the train at the peaks of those pulses that protrude from the range formed by the fences~$\alpha^{\pm}[k]$. This is illustrated in Fig.~\ref{fig:pulse counting}. When noise is negligible, and for appropriately chosen fences, $\hat{x}[k]$~given by~(\ref{eq:ccount sampling}) will be equivalent to the designed pulse train in the transmitter.

\begin{figure}[!b]
\centering{\includegraphics[width=8cm]{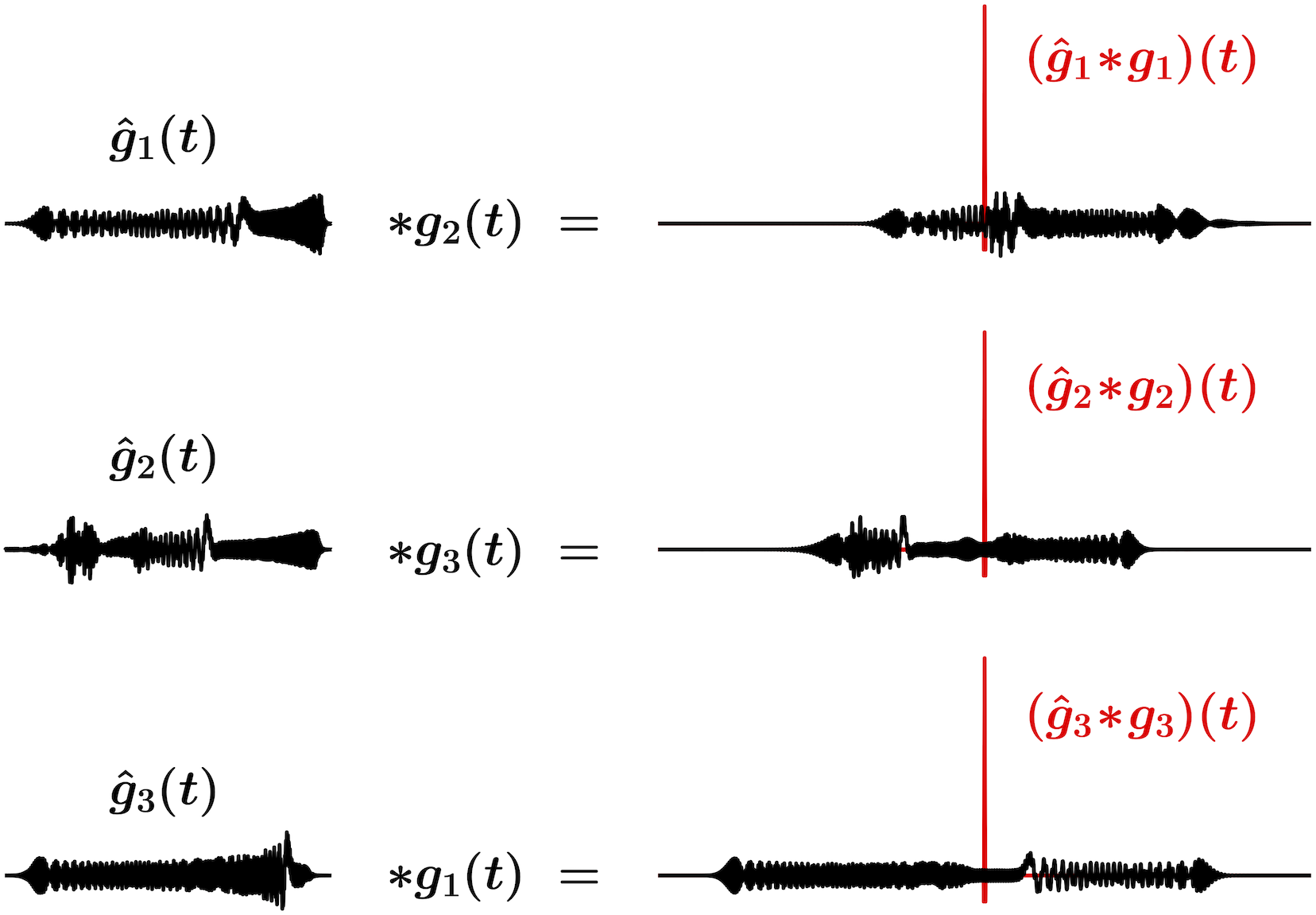}}
\caption{Example of autocorrelation (red) and cross-correlation responses for three filters~$\hat{g}_i(t)$ constructed by applying different allpass filters (consisting of 100~cascaded biquad sections) to RRC pulse.
\label{fig:autocross}}
\end{figure}
\addtocounter{figure}{1}
\begin{figure*}[!b]
\centering{\includegraphics[width=17cm]{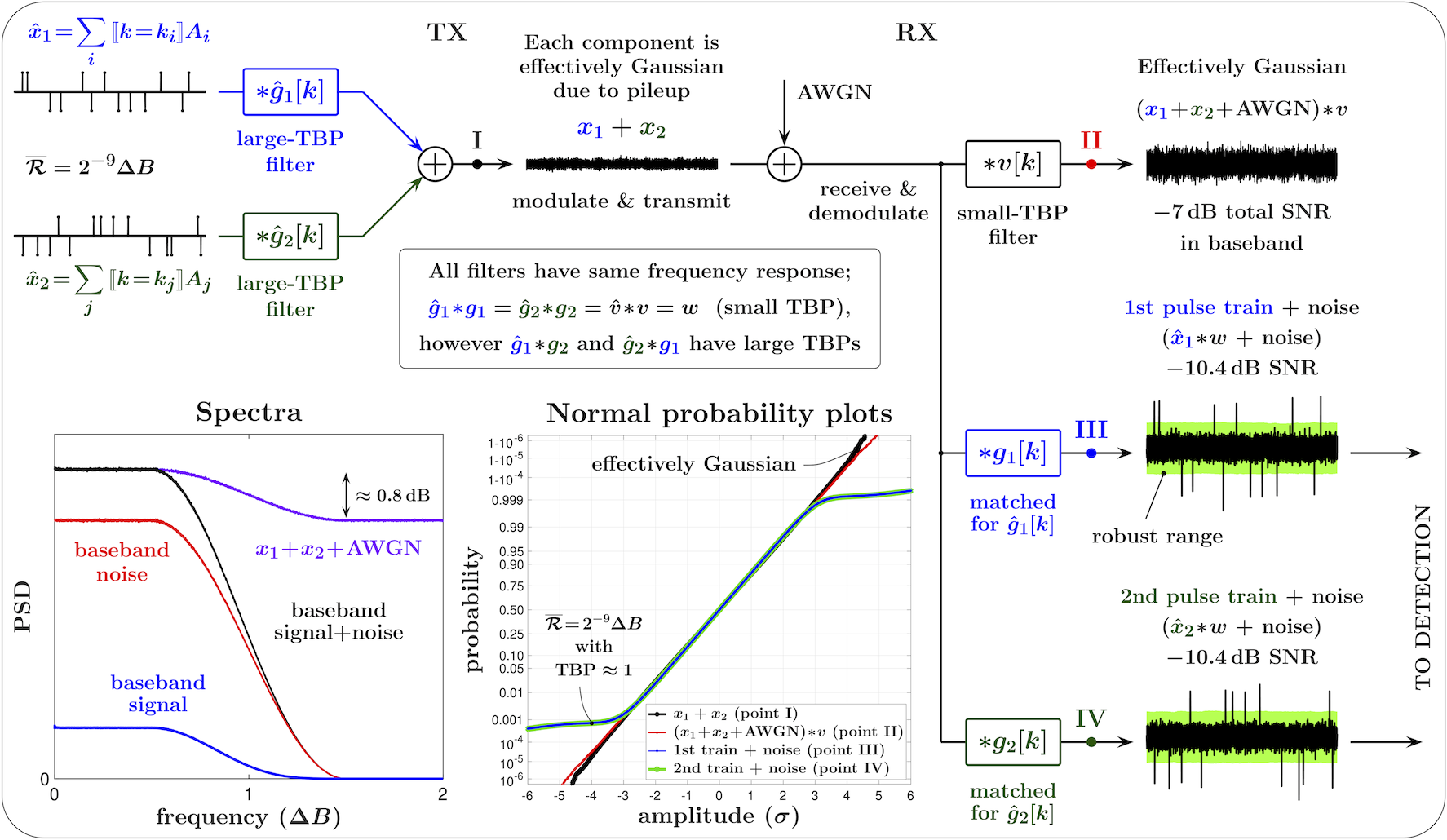}}
\caption{Illustrative example of using channel noise as cover signal for two-component payload.
\label{fig:TxRx1 detailed}}
\end{figure*}
\addtocounter{figure}{-2}
\begin{figure}[!t]
\centering{\includegraphics[width=8cm]{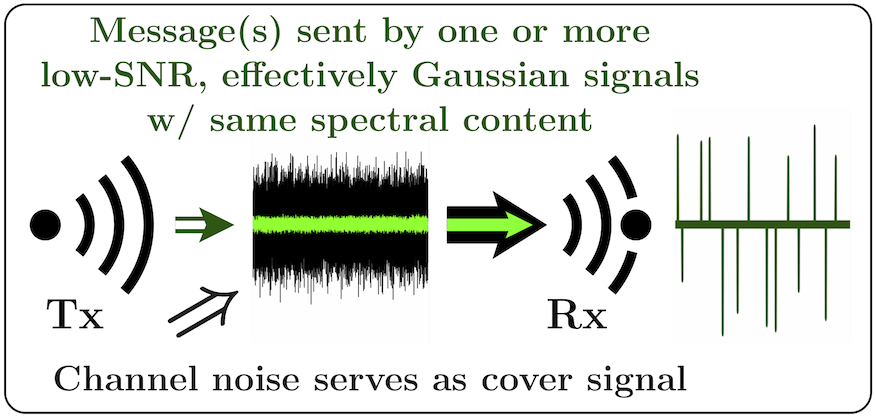}}
\caption{Using channel noise as cover signal.
\label{fig:TxRx1}}
\end{figure}
\addtocounter{figure}{2}
\begin{figure*}[!b]
\centering{\includegraphics[width=17.8cm]{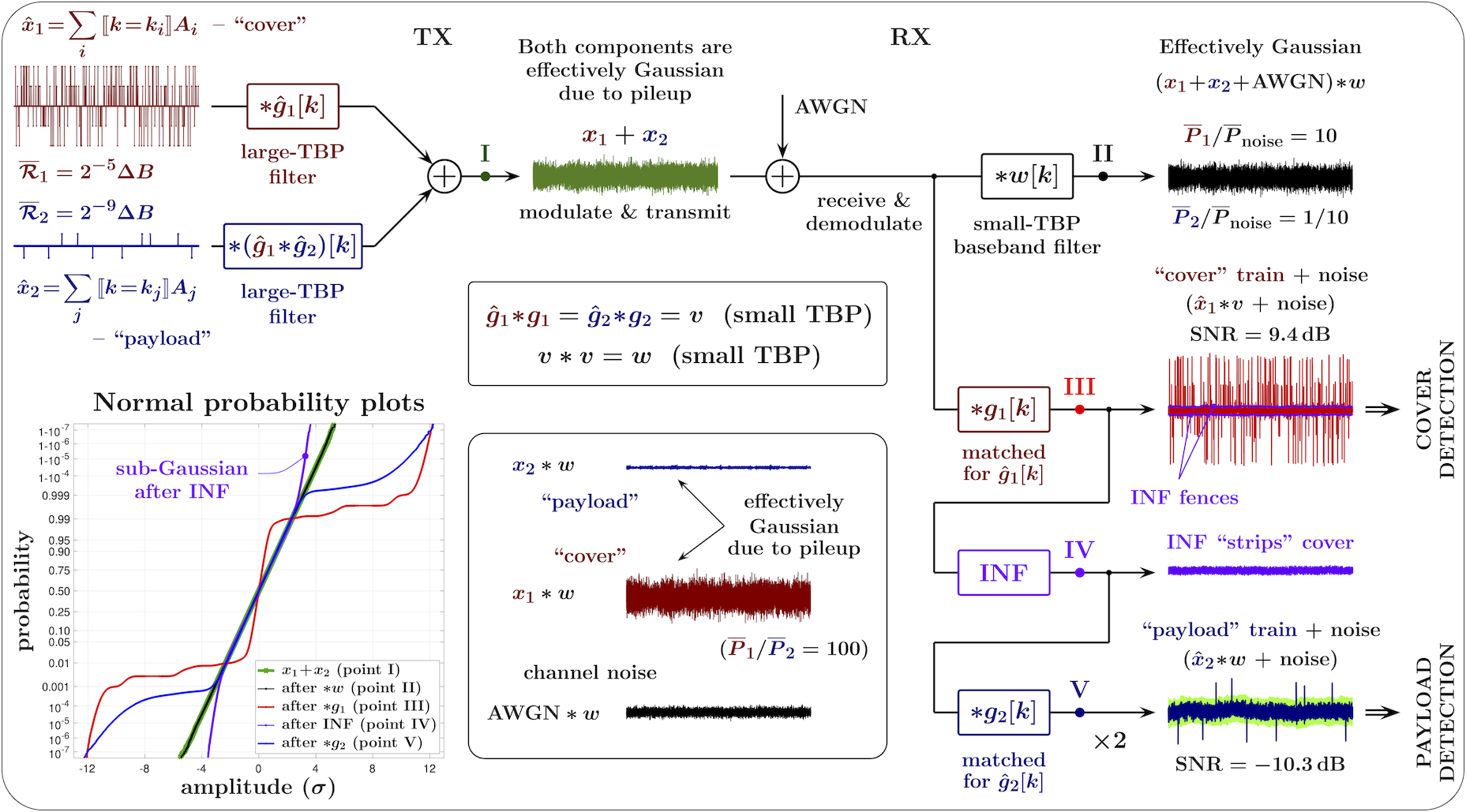}}
\caption{Illustrative example of adding INF-removable cover as outlined in Fig.~\ref{fig:TxRx2}. Power of cover exceeds power of payload by $\times 100$ (20\,dB), while both occupy effectively same spectral band.
\label{fig:TxRx2 detailed}}
\end{figure*}
\addtocounter{figure}{-2}
\begin{figure}[!t]
\centering{\includegraphics[width=7.6cm]{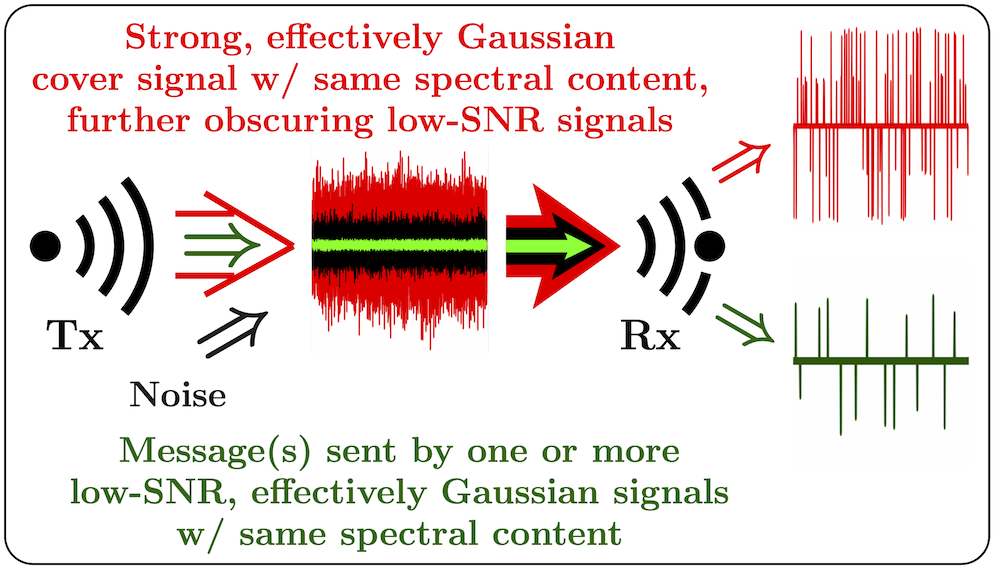}}
\caption{Additional INF-removable cover for low-SNR payloads.
\label{fig:TxRx2}}
\end{figure}

\section{Physical Layer Steganography} \label{sec:steganography}
To meet the undetectability requirement, in a steganographic system the stego signals should be statistically indistinguishable from the cover signals~\cite{Fridrich04searching, Fridrich05writing, Bash13limits, Bash15hiding}. For physical layer transmissions, this can perhaps be enhanced by requiring that the payload and the cover have the same bandwidth and spectral content, the same apparent temporal and amplitude structures, and that there are no explicit differences in the spectral and/or temporal allocations for the cover signals and the payload messages.

As discussed in Section~\ref{sec:ppileup}, by filtering a signal with a large-TBP filter one can ''mimic" the signal as effectively Gaussian, without affecting its spectral composition. Given a small-TBP filter~$v(t)$ with a particular frequency response, one can construct a great variety of filters~$\hat{g}_i(t)$ and~$g_i(t) = \hat{g}_i(-t)$ with the same frequency response yet much larger TBPs (e.g., orders of magnitude larger). Thus the filters~$\hat{g}_i(t)$ can be used in the transmitter to transform both the cover and the payload signals into effectively Gaussian, making them statistically indistinguishable from each other, while having the same spectral content. These filters can be constructed in such a way that their combined matched (autocorrelation) responses are equal to each other, $(\hat{g}_i\!\ast\!g_i)(t) = v(t)\!\ast\!v(-t) = w(t)$ for any~$i$, and have a small TPB, yet the convolutions of any $\hat{g}_i(t)$ with $g_j(t)$ for $i\ne j$ (cross-correlations) have large TBPs. Fig.~\ref{fig:autocross} illustrates this for three filters constructed by applying different allpass filters (consisting of 100~cascaded biquad sections) to an RRC pulse. Then a selected $i$-th signal component can be ``recovered" in the receiver by applying~$g_i(t)$, while the rest of the signal would remain effectively Gaussian.

For example, let us consider ${K\!\ge\! 2}$ effectively Gaussian waveforms ${x_i = (\hat{x}_i \!\ast\! \hat{g}_i)(t) = \sum_m A_m\, \hat{g}_i(t\!-\!t_m)}$. Then any $i$-th component in the sum ${x=\sum_{i=1}^K x_i}$ can be viewed as a ``payload," while the rest of the mixture can be considered a ``cover." To ``extract" the payload, we apply~$g_i(t)$ to this mixture:
\beginlabel{equation}{eq:mixture}
  (x \ast g_i)(t) = \sum_m A_m w(t\!-\!t_m) + \sum_j \Ibl j\ne i\Ibr\, (\hat{x}_j \!\ast\! g_{ji})(t)\,,
\end{equation}
where $g_{ji}(t)=(\hat{g}_j\ast g_i)(t)$. The right-hand side of~(\ref{eq:mixture}) can be treated as a pulse train affected by Gaussian noise. When the external noise is negligible (i.e. when the total SNR is much larger than $K$), for a mixture of equal-power waveforms the SNR for such a train will be ${\approx 1/(K\!-\!1)}$ (e.g. $-7\,$dB for a 6-component mixture).

\subsection{Using Channel Noise as Cover Signal} \label{subsec:channel noise}
The very existence of a detectable carrier (cover signal) may be a dead giveaway for the stego payload. For example, the mere presence of a sheet of paper implies the possibility of a message written in invisible ink. Therefore, the best steganography should be ``carrier-less," when the payload is covertly embedded into something ``ever-present." In the physical layer, such ``ideal" and unidentifiable cover signal is the channel noise. Such noise always includes the ever-present thermal noise as one of its components, and typically contains other (in general, non-Gaussian) natural and/or technogenic (man-made) components which make the noise non-white and characterized by time-variant parameters. Then, if the stego payload ``pretends" to be Gaussian, and its PSD everywhere is small enough to be well within the ``natural" variations in the PSD of the channel noise, any physically available band can be used to carry a virtually undetectable covert message. This approach is schematically illustrated in Fig.~\ref{fig:TxRx1}, where it is assumed that the messages are sent by pulse trains shaped with large-TBP filters.

In essence, all spread spectrum techniques for covert communications involve transmitting a signal that requires a limited bandwidth on a much wider bandwidth, thereby suppressing the PSD of the transmission below the noise floor~\cite{Bash13limits, Bash15hiding}. Even if the noise is stationary and accurate wideband measurements of its PSD $\mathcal{N}$ are available, detection of the mere presence of a signal with the peak PSD $\mathcal{S}\ll \mathcal{N}$ would require identifying an approximately $4.3\times \mathcal{S}/\mathcal{N} \lesssim 1\,$dB small ``bump" in the measured noise PSD during the transmission. (This task rapidly becomes more challenging for intermittent, short-duration transmissions, and/or for non-stationary noise.) Note, however, that even when the correct de-scrambling filter is applied to a pulse train shaped with a large-TBP filter, it does not change the average PSD of the transmitted signal, and so spectral measurements alone would still be insufficient for detection of the covert transmission. In contrast, for example, applying a known spreading sequence to a DSSS signal converts a wideband low-PSD signal into a narrowband high-PSD peak, and therefore enables its detection through spectral measurements.

As outlined in Fig.~\ref{fig:TxRx1}, the covert messages can be sent by several same-bandwidth pulse trains, each shaped with its own large-TBP filter, and Fig.~\ref{fig:TxRx1 detailed} provides a detailed illustrative example for a two-component mixture. For ${K\!\ge\! 1}$ equal-power components and dominant noise (e.g. when the noise power is much larger than the power of ${K-1}$ components), the SNR per component will be ${\approx \mathrm{SNR}/K}$. Thus the available pulse rate for the total SNR can be equally shared among $K$~low-SNR payloads, each requiring its own ``key" to extract the information.

\subsection{Adding INF-Removable Cover for Low-SNR Payloads} \label{subsec:obfuscation}
For a stego pulse train with a given rate, further increasing the power of the channel noise (say, by 10\,dB) can make the pulse train undetectable. For example, when the pulse rate is higher than the Shannon limit for the given SNR, neither synchronous nor asynchronous detection would be possible. However, such increase in the channel noise power can be accomplished by an additional pulse train, simply disguised as Gaussian. Then an INF in the receiver, in combination with the respective ``de-scrambling" filter, can effectively remove this additional noise, enabling the detection of the low-power payload. In addition, the higher-power pulse train can itself carry a lower-security (or decoy) message, and/or the timing information that enables synchronous pulse detection in the stego pulse train. Recovering this information from the ``extra cover" signal would still require knowledge of the respective scrambling filter used by the transmitter. This concept is schematically illustrated in Fig.~\ref{fig:TxRx2}, and Fig.~\ref{fig:TxRx2 detailed} provides its detailed illustrative example. Here, the power of the cover exceeds the power of the payload by a factor of~100 (by 20\,dB), while both occupy effectively the same spectral band. Note that even after the effective removal of the higher-SNR pulse train from the mixture by the INF, the stego message is still Gaussian, and still hidden behind the channel noise (and the remainder of the decoy/timing/``extra cover" signal). Thus its recovery still requires knowledge of the second scrambling filter ($\hat{g}_2[k])$) used by the transmitter.

\addtocounter{figure}{1}
\begin{figure}[!t]
\centering{\includegraphics[width=8.6cm]{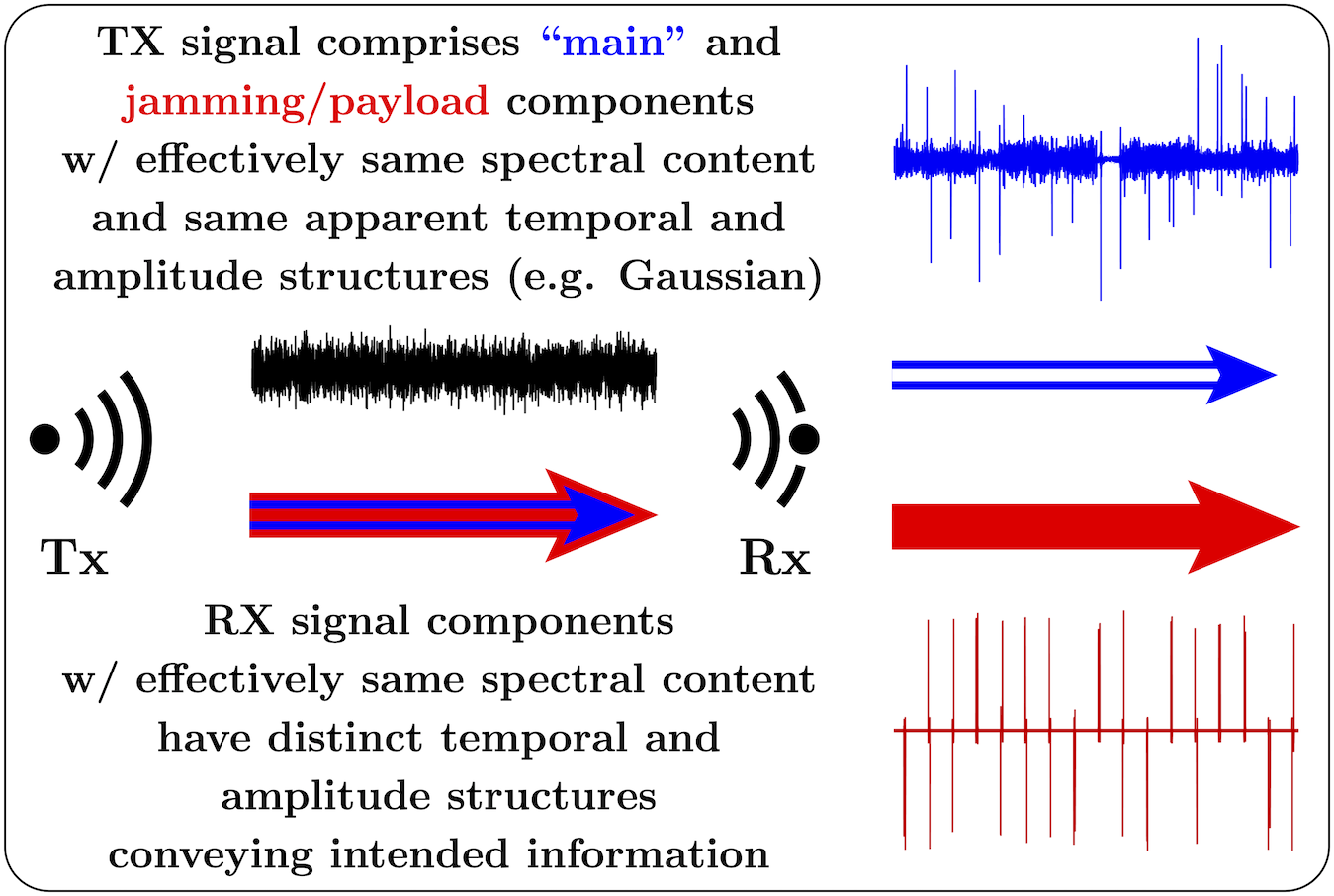}}
\caption{Basic concept of ``friendly in-band jamming."
\label{fig:friendly jamming concept}}
\end{figure}

\subsection{Friendly In-Band Jamming} \label{subsec:friendly jamming}
In our third example, the main message is transmitted using one of the existing communication protocols, but its temporal and amplitude structure is obscured by employing a large-TBP filter in the transmitter, e.g., made to be effectively Gaussian. This alone provides a certain level of security, since the intersymbol interference becomes excessively large and the signal cannot be recovered in the receiver without the knowledge of the scrambling filter. In addition, a jamming pulse train, disguised as Gaussian by another (and different) large-TBP filter, is added to the main signal. This jamming signal has effectively the same spectral content as the main signal, and its power is sufficiently large so that the main signal is unrecoverable even if the first scrambling filter is known. In the receiver, the jamming pulse train is removed from the mixture (and recovered, if it itself contains information), enabling the subsequent recovery of the main message. This concept is schematically illustrated in Fig.~\ref{fig:friendly jamming concept}.

\begin{figure}[!b]
\centering{\includegraphics[width=8.6cm]{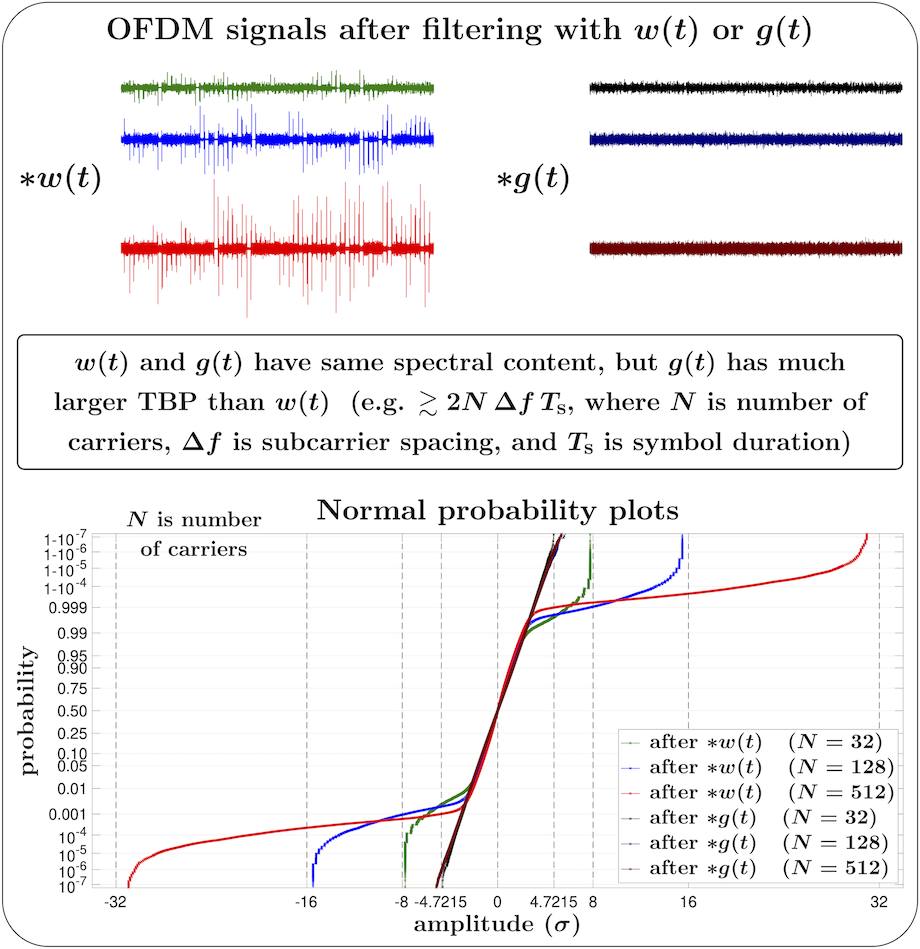}}
\caption{OFDM PAPR reduction by large-TBP filtering.
\label{fig:OFDM PAPR}}
\end{figure}
\begin{figure*}[!t]
\centering{\includegraphics[width=17cm]{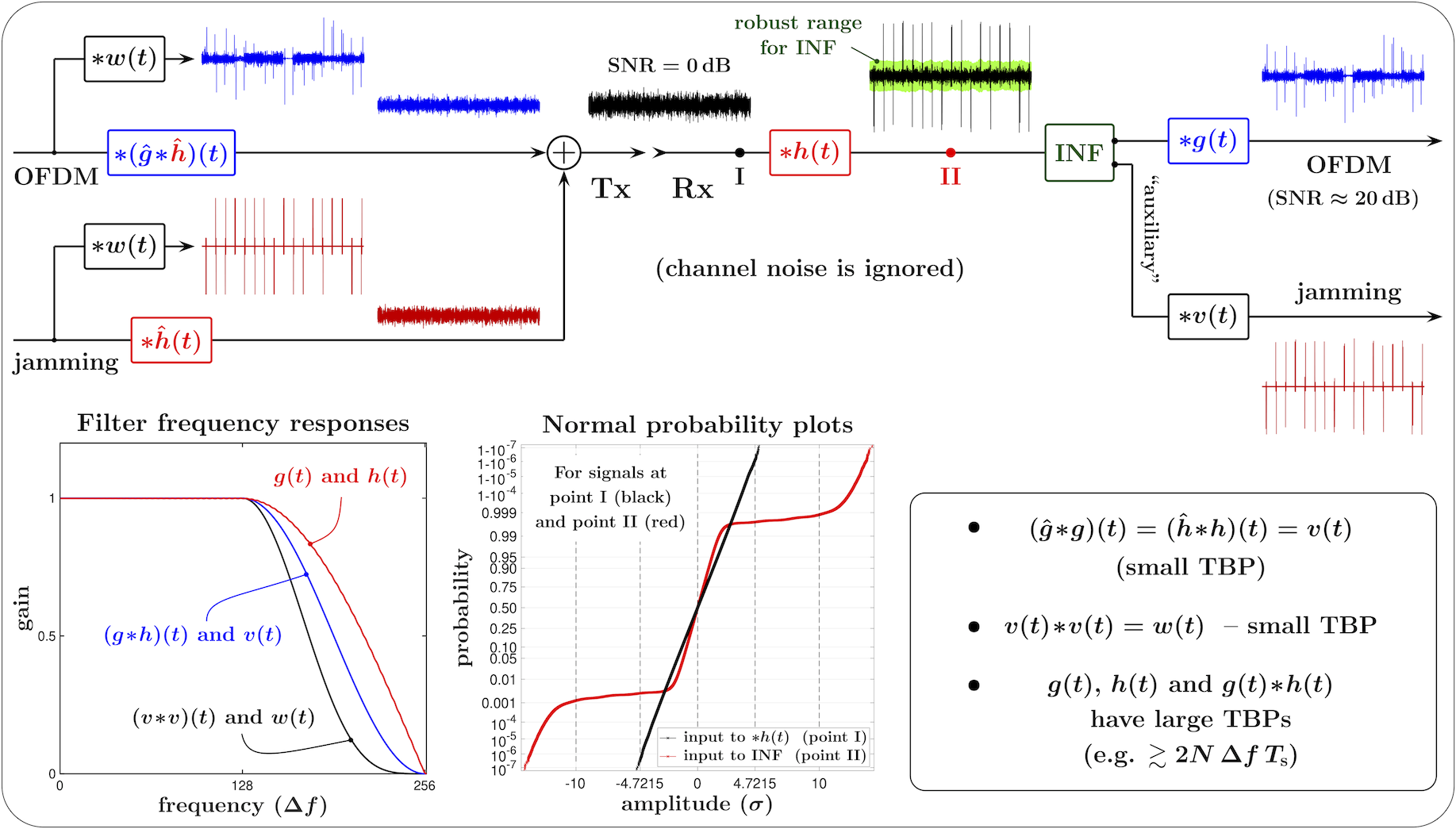}}
\caption{{\bf Friendly in-band jamming of OFDM signal:}~ Combination of linear and nonlinear filtering in receiver is used for effective separation of OFDM and ``friendly jamming" signals, although both signals in received mixture have effectively same spectral characteristics and temporal and amplitude structures, and there are no explicit differences in their temporal allocations.
\label{fig:OFDM jamming}}
\end{figure*}

\subsubsection{OFDM PAPR Reduction} \label{subsubsec:PAPR}
In addition to improved security, applying a large-TBP filter to the main signal reduces PAPR of large-crest-factor signals such as those in orthogonal frequency-division multiplexing (OFDM), as illustrated in Fig.~\ref{fig:OFDM PAPR}. Here, the simulated OFDM signals are generated without restrictions of the proportion of ``ones" and ``zeros" in a symbol, and thus they have the maximum achievable PAPRs (i.e. $2N$, where~$N$ is the number of carriers).

\subsubsection{Illustrative Example} \label{subsubsec:walk-through example}
In Fig.~\ref{fig:OFDM jamming}, the main signal is a high-PAPR OFDM signal, and the jamming signal is a high-PAPR impulse train with the spectral content in effectively the same band (see the frequency responses of the filters in the lower left panel of Fig.~\ref{fig:OFDM jamming}). After the filtering with large-TBP filters $(\hat{g}\!\ast\!\hat{h})(t)$ and $\hat{h}(t)$, respectively, both the OFDM and the jamming signals become effectively Gaussian, and so does their mixture that is being transmitted and received (see the black line in the normal probability plots shown in the lower middle panel of Fig.~\ref{fig:OFDM jamming}). (In this example, the channel noise is assumed to be relatively small and is not shown.) However, applying a filter~$h(t)$ matched for~$\hat{h}(t)$ in the receiver restores the high-PAPR structure of the jamming signal (see the red line in the normal probability plots), while the OFDM component remains Gaussian. Subsequently, the INF accomplishes both the mitigation of the jamming pulse train affecting the OFDM component and the extraction of the jamming signal. Applying the filter~$g(t)$ to the prime INF output effectively restores the original high-PAPR OFDM signal. If desired, the jamming pulse train is restored by applying the filter~$v(t)$ to the auxiliary INF output. 

The main properties of the filters used in this example are listed in the lower right panel of Fig.~\ref{fig:OFDM jamming}, and their frequency responses are shown in the lower left panel of the figure.

\section{Conclusion} \label{sec:conclusion}
For a finite pulse rate, the ideal pulse train~$\hat{x}(t) = \sum_j A_j \delta(t\!-\!t_j)$ has infinite bandwidth. Similarly, the designed digital pulse sequence~$\hat{x}[k] = \sum_j \Ibl k\!=\!k_j\Ibr\, A_j$ occupies the full Nyquist range of frequencies. On the other hand, the information encoded in a low-rate pulse train can also be transmitted, with simple binary coding and modulation, utilizing much smaller bandwidth. This is why communication with pulse trains can be considered a spread-spectrum technique. In contrast to other such techniques, however, wideband pulse trains are constructed without actual physical ``spreading" of a narrowband signal, and the ``spreading factor" (i.e. the fraction of a given bandwidth that is utilized) can be managed by changes in the information rate and/or the bandwidth of the pulse shaping filter. The main advantage of this approach lies in retaining control over the temporal and amplitude structure of the pulsed waveform. Such control, especially when combined with nonlinear filtering techniques, enables the development of a large variety of low-SNR and covert communication configurations.

Somewhat simplistically, other spread-spectrum approaches can be viewed as modulating a wideband carrier by a narrowband signal in the transmitter, then recovering the narrowband signal in the receiver. For example, since multiplication is associative, the spreading sequence in DSSS can be applied first to the carrier, forming a wideband carrier signal; then the wideband carrier is modulated by the narrowband signal. In FHSS and CSS, the frequency range of the carrier also spans a wide range. In the receiver, the respective demodulation (now combined with despreading) is used to produce the information-carrying narrowband signal. Because of the lack of ``excess bandwidth" over that needed to carry the information (for the given modulation type), the statistical properties of such a narrowband signal can no longer be significantly changed by linear filtering that leaves the spectral composition of the signal intact. On the other hand, as we illustrate throughout the paper, the temporal and amplitude structures of wideband signals are easily managed, and distinct outliers in pulsed waveforms can be made appear, disappear, and reappear without affecting the signals' spectral content.

\begin{figure}[!b]
\centering{\includegraphics[width=8.6cm]{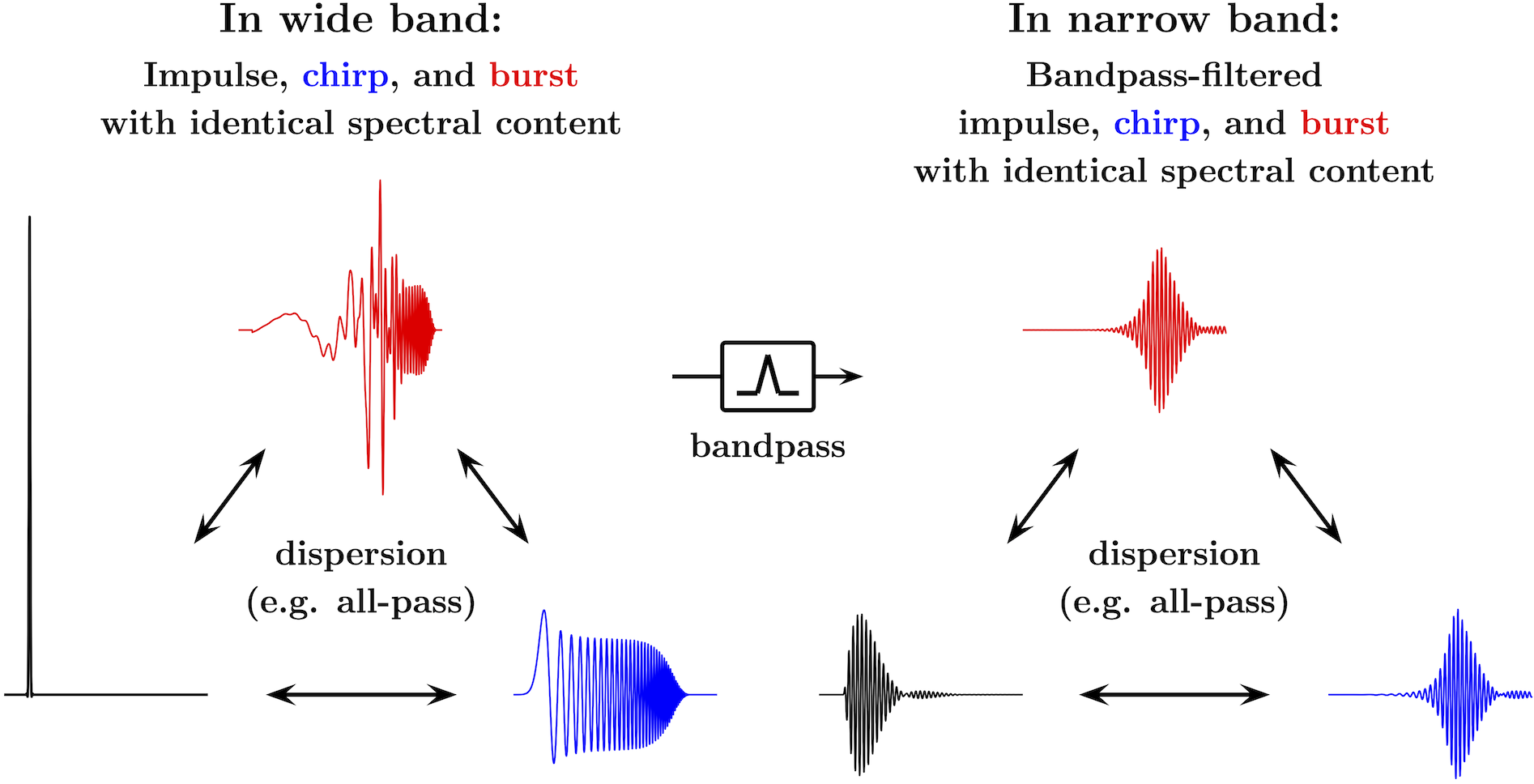}}
\caption{Effect of filtering on temporal and amplitude structure of signal is more apparent at wider bandwidth.
\label{fig:imp2chirp}}
\end{figure}

In general, the effect of filtering on the temporal and/or the amplitude structure of a signal is more apparent at wider bandwidths, as a broader frequency range results in finer time resolution. This is illustrated in Fig.~\ref{fig:imp2chirp}, where the impulse, the chirp, and the ``burst" signals have the same spectral content, and only the phases in their Fourier representations are different. These three signals can be morphed into each other by all-pass filtering that leaves their power spectral densities unmodified. In a wide band, such filtering drastically changes the time-domain appearance of the signal and its amplitude density, while in a narrow band (after the bandpass filtering) these changes are much less apparent and all signals maintain similar temporal and amplitude structures. Thus wideband pulsed waveforms extend our options for encoding low-rate information. Since wider bandwidth offers increased time resolution, the message can be encoded not only in the amplitudes of the pulses (e.g. as a binary sequence given by~(\ref{eq:ptrain equidistant})), but also in the pulse interarrival times, as represented by~(\ref{eq:ptrain}). This also adds to increased security, as various non-binary encoding protocols can be used for signals within the same physical band.

Further, control over the temporal and amplitude structures of wideband pulse trains carrying low-rate information provides for effective use of nonlinear filtering techniques. Such techniques can be employed for robust real-time asynchronous extraction of the information as well as for separation of signals with the same spectral content from each other. For example, the interquartile range can provide reliable measure for the average power of a Gaussian or sub-Gaussian signal, while being insensitive to changes in the power of such sparse signals as high-PAPR pulse trains. This enables the use of asynchronous detection (pulse counting) to extract both the amplitude and the timing information from randomized pulse trains. More generally, intermittently nonlinear filtering, when combined with waveform control by linear pulse shaping, allows effective separation of different signals with identical spectral profiles, facilitating development of versatile mixtures of same-band signals for covert and/or hard-to-intercept communications, and increasing robustness and quality of such communications in the presence of non-Gaussian noise. Two particular illustrative examples of using such mixtures are detailed in the previous section. 

\section*{Acknowledgment}
The authors would like to thank
Kendall Castor-Perry (aka The Filter Wizard);
James~E. Gilley of BK Technologies;
Jeff~E. Smith of Sierra Nevada Corporation;
Arlie Stonestreet\,\,II of Ultra Electronics ICE,
and Kyle~D. Tidball of Textron Aviation,
for their valuable suggestions and critical comments.
We extend our special gratitude to late Dr. Earl McCune (1956--2020), for encouraging us to address this topic. 


\end{document}